\documentclass[11pt,a4paper]{amsart}

\usepackage[a4paper,margin=3cm,footskip=.5cm]{geometry}

\usepackage{hyperref}
\usepackage{amsmath}
\usepackage{amssymb}
\usepackage{amsthm}
\usepackage{mathtools}
\usepackage{enumerate}
\usepackage{graphicx}
\usepackage{cite}
\usepackage{caption}
\usepackage{subcaption}
\usepackage{bm}
\usepackage{bbm}
\usepackage{float}
\usepackage{multirow}
\usepackage{makecell}
\usepackage{xcolor}
\usepackage{arydshln} 
\usepackage{caption}
\usepackage{boldline}
\usepackage{tabu}
\usepackage{array}
\usepackage{multirow}
\usepackage{booktabs}
\usepackage{hhline}
\usepackage{epstopdf}
\usepackage[pagewise]{lineno}\linenumbers

\newcolumntype{C}[1]{>{\centering\arraybackslash}m{#1}}

\newcommand{\EQ}{\begin{equation}}
\newcommand{\EN}{\end{equation}}
\newcommand{\EQS}{\begin{equation*}}
\newcommand{\ENS}{\end{equation*}}
\newcommand{\AL}{\begin{align*}}
\newcommand{\EAL}{\end{align*}}

\newcommand{\bea}{\bed\begin{array}{rl}}
\newcommand{\eea}{\end{array}\eed}

\newcommand{\rr}{{\hbox{{\rm I}{\kern -0.22em}{\rm R}}}}

\newcommand{\e}{\mathrm{e}}

\newcommand{\mi}{\mathrm{i}}

\newcommand{\D}{\mathrm{d}}
\newcommand{\Hes}{\mathrm{Hes}}
\newcommand{\Kou}{\mathrm{Kou}}

\usepackage{color}
\usepackage{amsaddr}
\usepackage{bbm}
\usepackage{bm}

\theoremstyle{remark}
\newtheorem{remark}{Remark}

\theoremstyle{plain}

\theoremstyle{plain}

\def\({\left(}
\def\){\right)}

\newcolumntype{L}[1]{>{\raggedright\let\newline\\\arraybackslash\hspace{0pt}}m{#1}}
\newcolumntype{M}[1]{>{\centering\let\newline\\\arraybackslash\hspace{0pt}}m{#1}}

\begin{document}

\nolinenumbers

\title[Stochastic Volatility and Double Exponential Jumps]{Calibration and Option Pricing with stochastic volatility and double exponential jumps}

\author{Gaetano Agazzotti }
\address{Soci\'{e}t\'{e} G\'{e}n\'{e}rale \& \'Ecole des Mines de Nancy}
\email{gagazzotti@crans.org}

\author{Claudio Aglieri Rinella}
\address{Soci\'{e}t\'{e} G\'{e}n\'{e}rale}
\email{claudio.aglieri-rinella@sgcib.com}

\author{Jean-Philippe Aguilar}
\address{Corresponding Author, Soci\'{e}t\'{e} G\'{e}n\'{e}rale}
\email{jean-philippe.aguilar@sgcib.com}

\author{J. Lars Kirkby}
\address{Georgia Institute of Technology}
\email{jkirkby3@gatech.edu}




\keywords{Calibration, Option Pricing, Heston model, Jump diffusion, Stochastic Volatility Models, Exotic Options, Volatility Surface, Double exponential distribution}


%
\begin{abstract}
This work examines a stochastic volatility model with double-exponential jumps in the context of option pricing. The model has been considered in previous research articles, but no thorough analysis has been conducted to study its quality of calibration and pricing capabilities thus far.
We provide evidence that this model outperforms challenger models possessing similar features (stochastic volatility and jumps), especially in the fit of the short term implied volatility smile, and that it is particularly tractable for the pricing of exotic options from different generations. The article utilizes Fourier pricing techniques (the PROJ method and its refinements) for different types of claims and several generations of exotics (Asian options, cliquets, barrier options, and options on realized variance), and all source codes are made publicly available to facilitate adoption and future research. The results indicate that this model is highly promising, thanks to the asymmetry of the jumps distribution allowing it to capture richer dynamics than a normal jump size distribution. The parameters all have meaningful econometrics interpretations that are important for adoption by risk-managers.
\end{abstract}

\maketitle

\section{Introduction}\label{sec:intro}


Stochastic volatility (SV) models address two of the primary stylized facts of financial markets, namely volatility clustering, which is characterized by periods of increased market fluctuations and relative stability, as well as heavy-tailed asset returns, see \cite{cont2001empirical, cont2007volatility} among many other references. A prominent example within the SV model family is the Heston model \cite{heston1993closed}, that enjoys a broad popularity due notably to its amenability to efficient option pricing using Fourier transform techniques, see for instance \cite{corsaro2019general,unifiedBarrierBermudanStoKirkby,ballotta2022smiles,mackay2023analysis,zhong2023valuation,perotti2024pricing}; see also \cite{LittleHestonTrap} for an appropriate choice of the characteristic function to implement.


\subsection{Calibration and option pricing with SV and jumps}

A primary goal of quantitative modeling is to devise a model that not only captures observed market phenomena but also ensures reliable pricing for a variety of financial products, especially in the context of exotic option pricing.
In the foreign exchange markets, light exotic derivatives including barrier options are so prevalent and liquid that joint calibration of these exotics with the vanilla market is both feasible and beneficial for capturing the market's risk-neutral measure \cite{carr2010class}, assuming sufficient model richness. In markets with liquid volatility and variance derivatives trading, such as the S\&P500 and its associated VIX options, joint calibration is of particular interest \cite{guyon2022vix,XuAlexSVIX} to enable pricing of variance sensitive exotics.  In general, the key to a successful calibration is finding a model that is sufficiently rich to capture the vanilla market (and potentially some exotic markets concurrently), and is suitable for pricing the targeted exotic contracts. Moreover, as calibration is a computationally expensive process, the model should ideally lend itself to efficient pricing to be practically useful.

As with other model families such as
L\'evy processes (and in particular jump-diffusions), SV models offer various efficient pricing algorithms that make them ideal for calibration and pricing of certain exotic options. Moreover, SV models address a well-known limitation of L\'evy models, which is their tendency for the model implied volatilities to flatten out at larger maturities, due to the iid assumption of L\'evy return increments. 
The trade-off is that they can struggle to simultaneously capture both a steep short expiry skew with a longer term persistent skew in the volatility surface. The presence of steep skews for short-expiry options is often attributed to the possibility of large jumps in the underlying price, which can occur during an arbitrarily small time interval, for instance as a consequence of the reaction/overreaction of markets to good and bad news \cite{Overreaction}.
In this respect, L\'evy models have an advantage in capturing steep volatility skews at short maturities, which motivates the introduction of a jump component in the stochastic volatility models. Stochastic Volatility Jump Diffusions (SVJ) combine the advantages of each model, and often calibrate well in equity and FX markets. The additional parameters typically produce better calibrations, but at the cost of more complicated simulation and derivatives pricing.

The incorporation of a jump component in a stochastic volatility model dates back to the work of \cite{bates1996jumps}, the ``Bates model", which augments the Heston stochastic volatility model with log-normally distributed jumps. Not only does the jump component help to capture short expiry skews, but it ``frees up" the parameters of the stochastic volatility process to focus on modeling the volatility term structure and forward skew, which is important for certain exotic options like forward-starts and cliquets / equity indexed annuities (EIAs) \cite{marabel2014pricing,cui2017equity,kirkby2023valuation,ai2023valuing}. SVJ models are thus able to capture the prominent features of a volatility surface and retain the advantages of an interpretable stochastic volatility component, which is advantageous for risk management.
Since the seminal work of \cite{bates1996jumps}, SVJ models have become prominent in financial modeling, with recent research including \cite{cheang2020representation,kirkby2020efficient,bosserhoff2023robustness,ma2020efficient}. \\
While our focus in this work is on SVJ, other variations on stochastic volatility worth mentioning include stochastic local volatility models (SLV) \cite{ogetbil2023extensions,cui2018general,ma2024vix,van2020collocating}, as well as the more recent rough volatility models  \cite{gassiat2023weak,jacquier2023deep,richard2023discrete,guyon2022vix,jacquier2023deep} and rough stochastic local volatility \cite{yang2024general}, which have shown great promise in certain applications, including joint SPX/VIX calibration. This research avenue is promising as recent empirical studies indicate that these models are able to reproduce several stylized facts for realized and implied volatilities \cite{Gatheral1, Fukasawa1, BRANDI1}. 
Some authors also consider decomposing the volatility into two stochastic processes: the empirical results in \cite{Christoffersen_et_al} demonstrate the improvement achieved by incorporating two-factor dynamics into the classical Heston model, while in \cite{KhatibYoussef} a study on a double-factor SVJ was conducted; \cite{ChangYingWangFract}, instead, studies the case where one of the two volatility processes is driven by a fractional Brownian motion. Another possible generalization is to add a jump component in the volatility process: \cite{SunLiuGuo, SeneNdeye} introduce such a model with respectively asynchronous and synchronous jump component of the log-price process.

\subsection{The HKDE model}
This work studies a particular variation of the Heston model that combines it with a double-exponential jump distribution for the spot process introduced in \cite{kou2002jump}, which we will refer to as the Heston-Kou Double Exponential (HKDE) model.  
We also note that some works have combined stochastic volatility with double-exponential distribution of jumps and other customizations in order to capture specific features. For instance, stochastic rates have been incorporated in the HKDE model in \cite{Chen2017}; \cite{ChangYingWang} also extends it by allowing jump intensity to be stochastic. In this work, we contribute to the strain of research on SVJ modeling,  and in particular on the Heston model with the double exponential jump distribution introduced by Kou in \cite{kou2002jump}. This model has been selected for its ability to achieve high quality volatility surface fits with a limited number of model parameters. We will focus on the calibration aspects of this model and the subsequent pricing of exotic options.

For risk-managing exotic options, interpretability and stability of the calibrated model parameters are often as important as the quality of fit to the vanilla market. The HKDE model comprises nine parameters. Although this is not minimal, it is fewer than those in other sophisticated models (for instance, the model by \cite{ChangYingWang} includes seventeen parameters). Moreover, each parameter of HKDE represents an economically meaningful quantity that can be further estimated through econometric techniques.  Surprisingly, only a few works consider the simpler HKDE model, and no thorough study has been conducted to test its goodness of fit (notably in terms of volatility surface fitting), nor has it been tested for pricing advanced generation of exotics (only \cite{ZhangSumei} introduces HKDE to price one specific type of contracts - forward starting options-  with the COS method). The objective of this paper is therefore to bridge this gap and provide an in-depth analysis of the HKDE model in terms of calibration, performance, and pricing.

\subsection{Key contributions}
This work provides several contributions to the literature on derivatives pricing and model calibration, summarized as follows:
\begin{itemize}
    \item Calibrating the HKDE model on single stock option data, demonstrating a better fit of the volatility smile when compared to challenger models featuring stochastic volatility and / or jumps (Heston, Bates, Bilateral Gamma Motion);
    \item Providing a deep investigation of the smile sensitivity to the HKDE model parameters to further the understanding/interpretability of the model parameters;
    \item Pricing and analysis for several generations of exotic contracts (Asian, Discrete Variance Options, Cliquet, Barrier).
\end{itemize}
Moreover, for the reader's convenience, Python source code for model calibration and exotic option pricing is made publicly available at \url{https://github.com/jkirkby3/fypy}.

The paper is organized as follows: Section \ref{sec:model} introduces the HKDE model and related quantities, as well as a detailed sensitivity analysis of its parameters. Section \ref{sect:SurfacePriceCali} details the calibration methodology. In Section \ref{sec:MarketCalibrationResults} we conduct a market calibration for four equity assets (Amazon, Shopify, Spotify and Netflix); the goodness of fit is compared to challenger models via different error metrics (MAPE and RMSE). In Section \ref{sec:Exotics}, we compute the prices of several exotic derivatives, utilizing refinements of Fourier pricing techniques. Section \ref{sec:Conclu} is dedicated to conclusions. Last, we have included two Appendices about the PROJ method: Appendix A provides technical details on the derivation of the projection coefficients, while Appendix B discusses the performance of this method compared to Monte Carlo pricing, providing justification for our choice of the PROJ method for calibration and exotic option pricing in this paper.

\section{Model definition and properties}\label{sec:model}

In all of the following, we let $(\Omega, \mathcal F, \{\mathcal F_t\}_{t\geq 0},  \mathbb Q)$ be a risk-neutral probability space equipped with its natural filtration. Unless otherwise mentioned, all expectations are considered under $\mathbb{Q}$, the risk-neutral measure.

\subsection{Model specification}

We define the HKDE model by adding a specific jump-diffusion component to the usual Heston dynamics. More precisely, we assume that, under $\mathbb Q$, we have
\EQ
\label{eq:model_dynamics}
\begin{cases}
    \D S_t = \mu S_t \D t + \sqrt{V_t} S_t \D W^1_t + S_t\left(e^{J_t}-1\right)\D N_t~,\\
    \D V_t = \kappa\left(\theta-V_t\right)\D t + \sigma_v \sqrt{V_t}\D W^2_t
    ,
\end{cases}
\EN
where the processes $W^1=(W^1_t)_{t\geq 0}$ and $W^2=(W^2_t)_{t\geq 0}$ are two correlated Brownian motions, \textit{i.e.} $\D\langle W^1,W^2\rangle_t = \rho \D t$ with $\rho\in[-1,1]$. The process $J := (J_t)_{t\geq 0}$, which models the log-size of relative price jumps, is characterized by a double-exponential density defined over the entire real line by
\EQ
\label{eq:jump_density}
f_J(y) = p\eta_1e^{-\eta_1 y }\mathbbm{1}_{\{y\geq 0\}}+(1-p)\eta_2e^{-\eta_2 y }\mathbbm{1}_{\{y< 0\}}
, 
\quad
p\in[0,1]\hspace{1.5mm},\hspace{1.5mm} \eta_1,\eta_2> 0
.
\EN
The real number $p$ (\textit{resp.} $1-p$) represents the probability of jumping upward (\textit{resp.} downward); conditionally on jumping upward (\textit{resp.} downward), the size of jump follows an exponential distribution of parameter $\eta_1$ (\textit{resp.} $\eta_2$). Note that the asymmetry induced by $\eta_1, \eta_2$ and $p$ allows HKDE to capture more complex phenomena than Bates (where the jump size distribution is symmetric). Not only is this preferred from an empirical perspective (matching the asymmetry generally found in asset price jumps), but it provides additional flexibility to the model when calibrating to the observed market for derivatives. Last, the process $N := (N_t)_{t\geq 0}$ is a Poisson process with rate $\lambda >0$, and the volatility process $V := (V_t)_{t\geq 0}$ is a CIR process with mean $\theta$, speed of mean-reversion $\kappa$ and volatility of volatility $\sigma_v$. \\

\subsection{Characteristic function}\label{subsec:char}

If we assume that the jump process $J$ is independent of the Wiener processes $W^1$ and $W^2$, then the characteristic function of the log price process in \eqref{eq:model_dynamics} satisfies
\begin{equation}\label{Phi_HKDE}
    \phi_{\mathrm{HKDE}}(\xi, t)
    :=
    \mathbb E \left[ e^{i\xi \ln S_t} \right]
    = \phi_{\mathrm{Hes}} (\xi,t) \phi_{\mathrm{Kou}} (\xi,t)
    ,
    \quad
    (t,\xi) \in \mathbb R_+ \times \mathbb R
    .
\end{equation}
The characteristic function of the Heston model is known to be
\begin{align*}
\quad\phi_\Hes(\xi,t) = &\exp{(i\xi(\log S_0 + (r-q)t)}\\
&\times\exp(\eta\kappa\sigma_v^{-2}((\kappa-i\xi\sigma_v\rho-d(\xi))t-2\log((1-g(\xi)e^{- d(\xi)t})/(1-g(\xi))) ))\\
&\times \exp(V_0^2\sigma_v^{-2}(\kappa-i\xi\sigma_v\rho-d(\xi))(1-e^{-d(\xi)t})/(1-g(\xi)e^{-d(\xi)t})),
\end{align*}
where $V_0$ is the variance at $t=0$, and the $d$ and $g$ functions are defined as
\begin{equation*}
\begin{cases}
    d(\xi)\coloneqq \sqrt{(i\xi\sigma_v-\kappa)^2+\sigma_v^2(i\xi+\xi^2)},\\
    \displaystyle g(\xi)\coloneqq \frac{\kappa-i\xi\sigma_v\rho-d(\xi)}{\kappa-i\xi\sigma_v\rho+d(\xi)}.
\end{cases}
\end{equation*}
Also, the characteristic function of the jump component is well known and can be written as
\EQ
\quad\phi_\Kou(\xi,t) = \exp\left( t  \left(\lambda \left(\frac{p\eta_1}{\eta_1-i\xi}+\frac{(1-p)\eta_2}{\eta_2+i\xi}-1\right)+\omega_{\mathrm{KDE}}\cdot i\xi \right) \right)
,
\EN
where the term $\omega_{\mathrm{KDE}}$ ensures that the combined model is risk-neutral, and is defined as follows:
\[
\omega_{\mathrm{KDE}}:= -\lambda \Big( \frac{p \eta_1}{\eta_1 -1}  + \frac{(1-p)\eta_2}{\eta_2 + 1} -1 \Big).
\]

\subsection{Cumulants}\label{subsec:cumulants}

The cumulant generating function is defined as
\begin{equation}
    K_{\mathrm{HKDE}}(p) 
    :=
    \ln \mathbb E \left[ e^{ p \ln S_t} \right]
    =
    \phi_{\mathrm{HKDE}} (-ip, t)
    ,
\end{equation}
and the $n^{th}$ cumulant is $\kappa_{
n}(t)=K_{\mathrm{HKDE}}^{(n)}(0,t)$. The cumulants of HKDE model can be decomposed as $\kappa_n(t) = \kappa_{\mathrm{Hes},n}(t)+\kappa_{\mathrm{Kou},n}(t)$ by leveraging the properties of the characteristic function. The cumulants $\kappa_{\mathrm{Hes},n}(t)$  are provided in \cite{fang2009novel}, and the first four of $\kappa_{\mathrm{Kou},n}$ are the following:
\begin{align*}
    \kappa_{\mathrm{Kou},1}(t) &= t\left(\lambda\left(\frac{p}{\eta_1}-\frac{1-p}{\eta_2}\right) +\omega_{\mathrm{KDE}}\right), \\
    \kappa_{\mathrm{Kou},n}(t) &= n!t\lambda\left(\frac{p}{\eta_1^n}+(-1)^n\frac{1-p}{\eta_2^n}\right),\quad\forall n\geq 2. \\
\end{align*}
Knowledge of the cumulants is particularly useful when pricing with certain Fourier methods (as below), which utilize these cumulants to set the boundaries of a numerical grid.


\subsection{Parameters Analysis}\label{sec:paramana}
In order to gain more intuition on the HKDE model, we show how the parameters influence the (spot) implied volatility surface. Specifically, we begin with a set of parameters $P = (V_0, \theta, \kappa, \sigma_v,\rho,\lambda, p, \eta_1,\eta_2)$, and independently modify each parameter of $P$ to observe the impact on the smile's level, shape, convexity, etc. 

For this analysis, we consider two distinct maturities, using the parameters detailed in Table \ref{table:CalibParams} for the underlying asset SHOP. The corresponding implied volatility curves are then plotted showing the standard measure of moneyness, $\ln \left(K/S_0\right)$. 

\subsubsection{Volatility parameters}

{
\begin{figure}
    \centering
    \begin{subfigure}{0.4\textwidth}
      \centering
      \includegraphics[width=\linewidth]{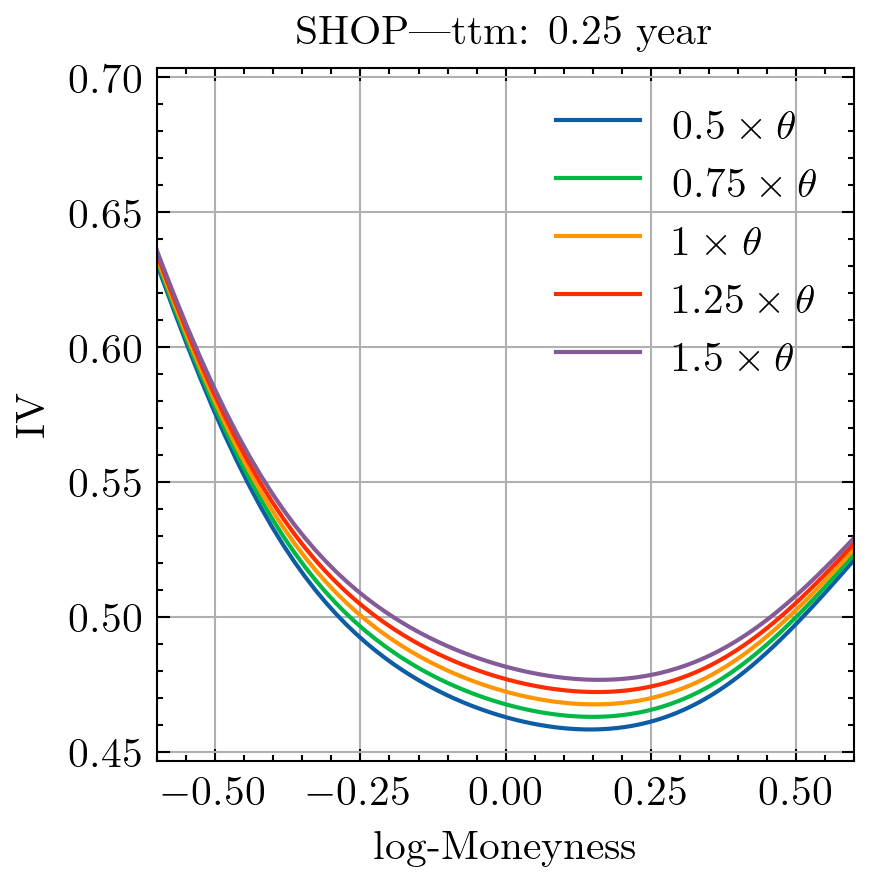}
      \caption{\footnotesize{Bumping $\theta$ - maturity 3 months.}}
    \end{subfigure}
    \begin{subfigure}{0.4\textwidth}
      \centering
      \includegraphics[width=\linewidth]{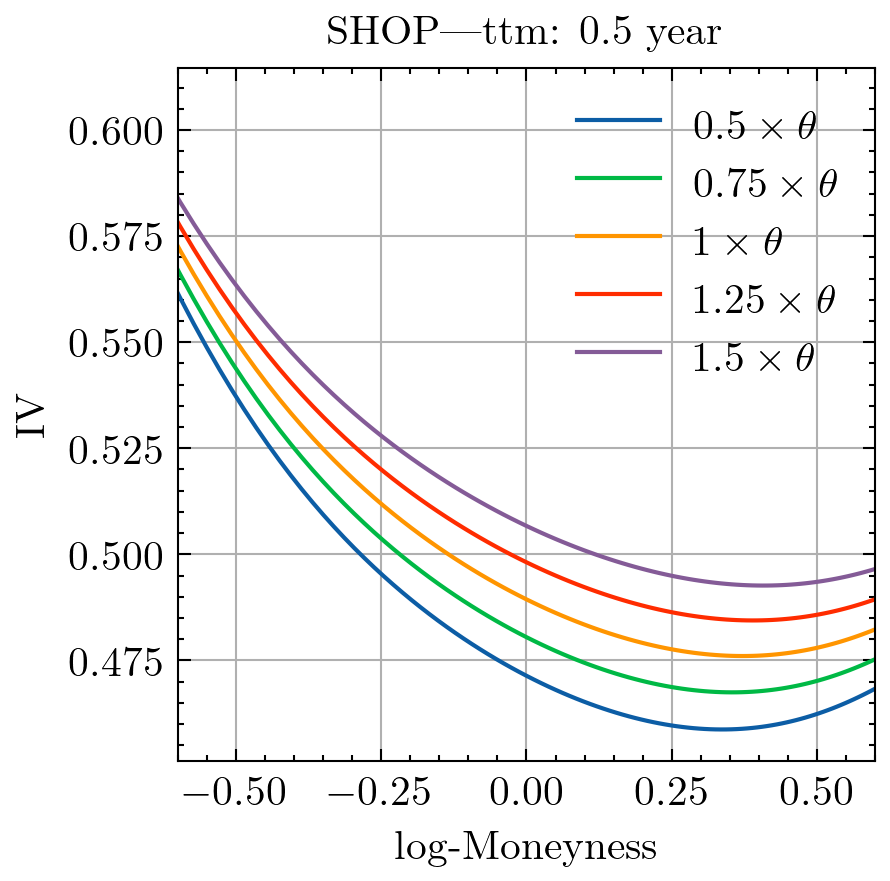}
      \caption{\footnotesize{Bumping $\theta$ - maturity 6 months.}}
    \end{subfigure}\vspace{0.5cm}
    \begin{subfigure}{0.4\textwidth}
      \centering
      \includegraphics[width=\linewidth]{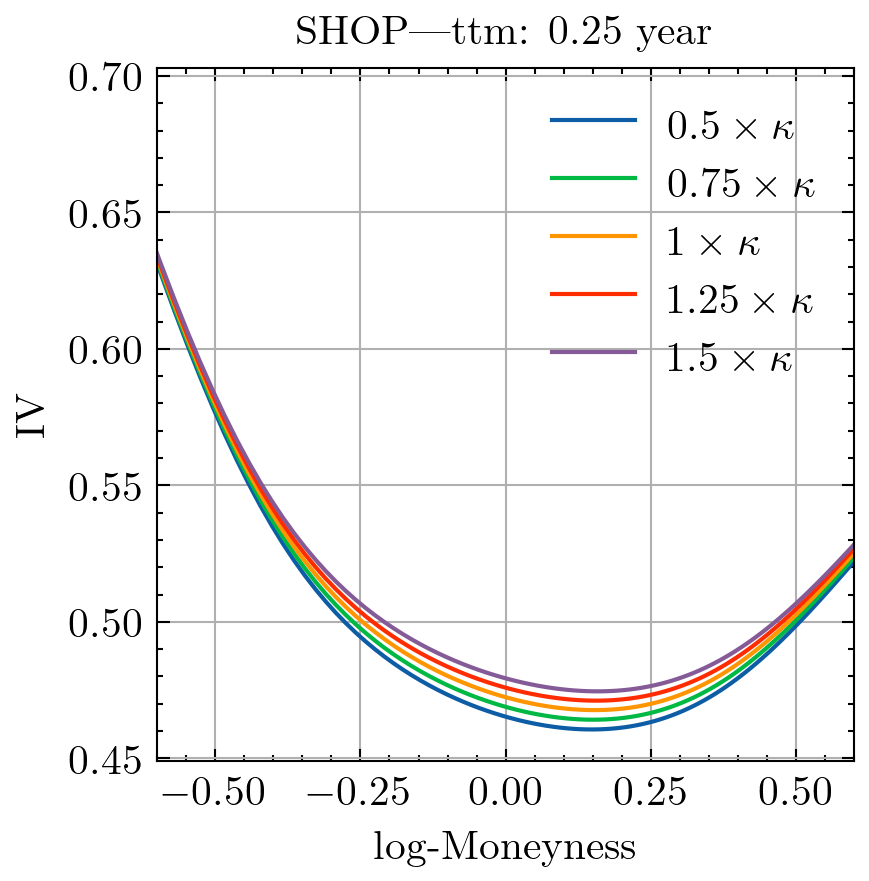}
      \caption{\footnotesize{Bumping $\kappa$ - maturity 3 months.}}
    \end{subfigure}
    \begin{subfigure}{0.4\textwidth}
      \centering
      \includegraphics[width=\linewidth]{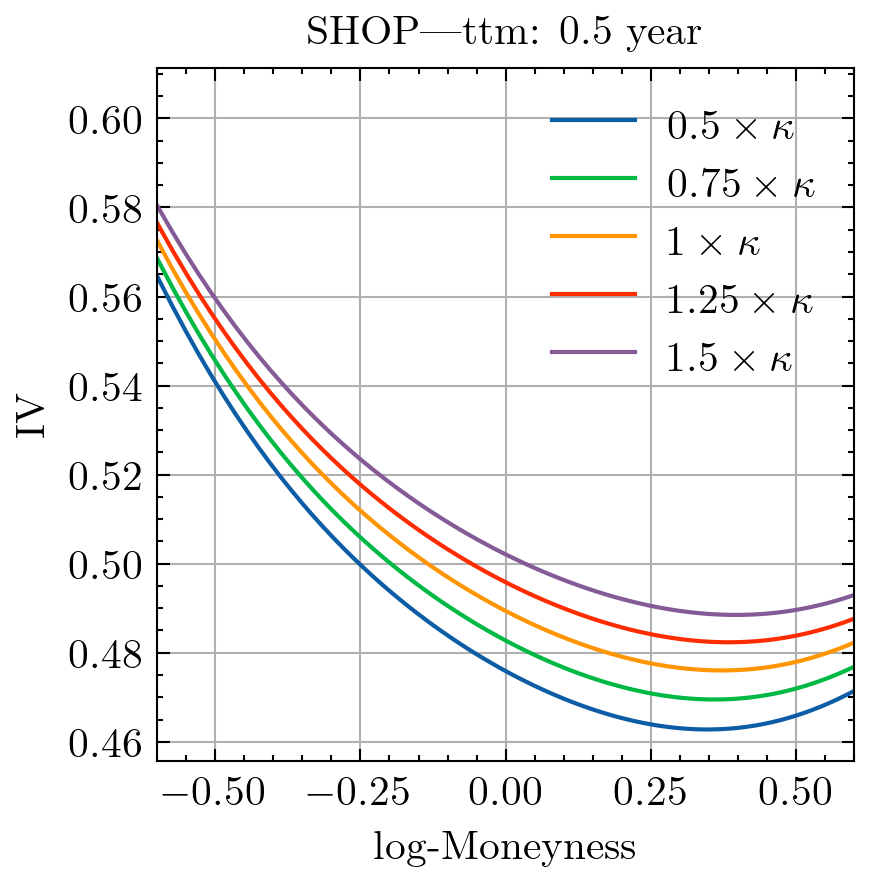}
        \caption{\footnotesize{Bumping $\kappa$ - maturity 6 months.}}   \end{subfigure}\\
    \caption{\color{black}
    \footnotesize{Effects of variations in the parameters $\theta$ and $\kappa$ on the implied volatility smiles. The plots represent implied volatility curves, for maturities of 3 and 6 months. 
    }
    }
\label{fig:bumping_mean_reversion}
\end{figure}
}

The first parameters we analyze are the components of the mean-reversion part of the volatility process, \textit{i.e.}, the mean $\theta$ and the speed of mean-reversion $\kappa$. The effects on volatility appear to be quite similar for both parameters $\theta$ and $\kappa$. More precisely, as shown in Figure \ref{fig:bumping_mean_reversion}, when the values of these parameters increase, the volatility smile shifts upward.  We note from Table \ref{table:CalibParams} that $\theta > V_0$, so increasing the speed of mean reversion to a higher level of variance will intuitively lift the volatility smile. Similarly, increasing the long term level of variance, $\theta$, tends to lift the level of volatility across the smile, with the greatest increase in the ATM region.  We also display in Figure \ref{fig:term_structure_kappa_theta} how modifying the $\theta$ and $\kappa$  parameters impacts the term structure of the ATM implied volatility (as a function of time to maturity). Given that $V_0 < \theta$, it is clear that increasing the long term level of  $\theta$ should have the greatest impact on the ATM level of longer dated maturities.  In particular, longer expiries allow more time for the volatility level to reach its higher long term level, and a larger percentage of the contract life will be spent at that level. A similar argument holds for $\kappa$.

Last, we note that an increase in the volatility of volatility $\sigma_v$ heightens the convexity of the smile and rotates the curve (see Figure \ref{fig:bumping_sigma_v}); the same behavior occurs for both considered maturities.  As $\sigma_v$ primarily controls the properties of the distribution tails --- governing the range of probable paths of volatility, and hence the likely dispersion in the underlying --- its influence affects both the skew and kurtosis of the log-price transition density. The skew impact is reflected via the rotation of the smile (the slope of which near the ATM point is also called the implied volatility skew), while the kurtosis manifests as an increase in smile convexity and a change in the level of the wings.

{\begin{figure}
    \centering
    \begin{subfigure}{0.4\textwidth}
      \centering
      \includegraphics[width=\linewidth]{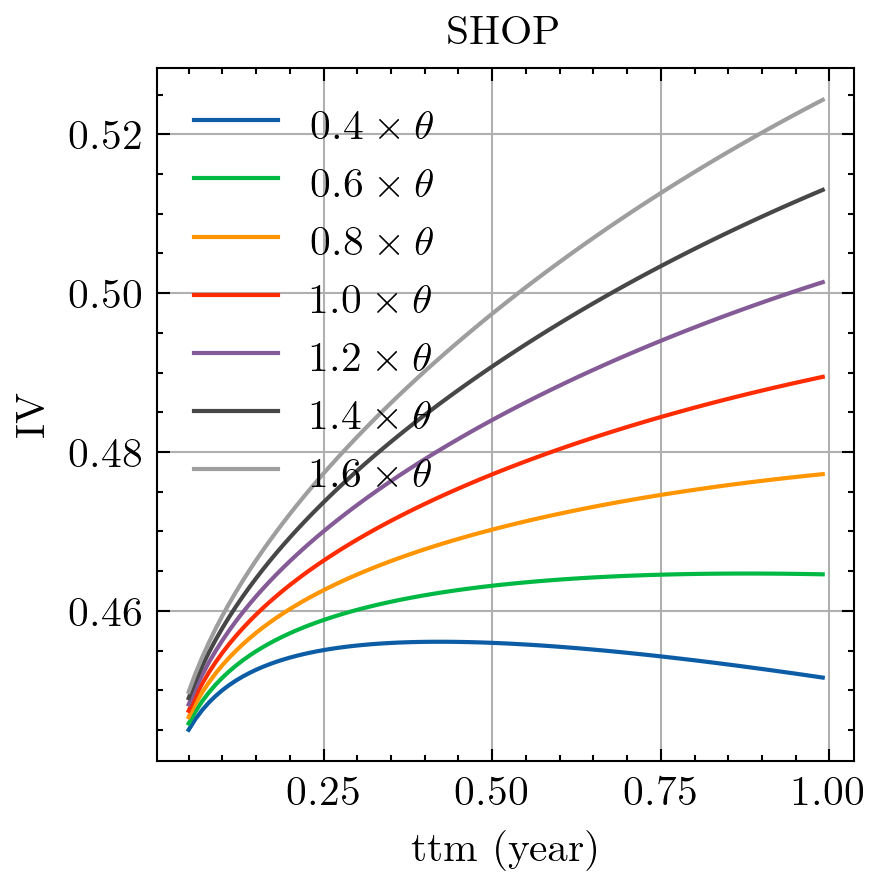}
      \caption{\footnotesize{Bumping $\kappa$ - Effects on the term structure.}}
    \end{subfigure}
    \begin{subfigure}{0.4\textwidth}
      \centering
      \includegraphics[width=\linewidth]{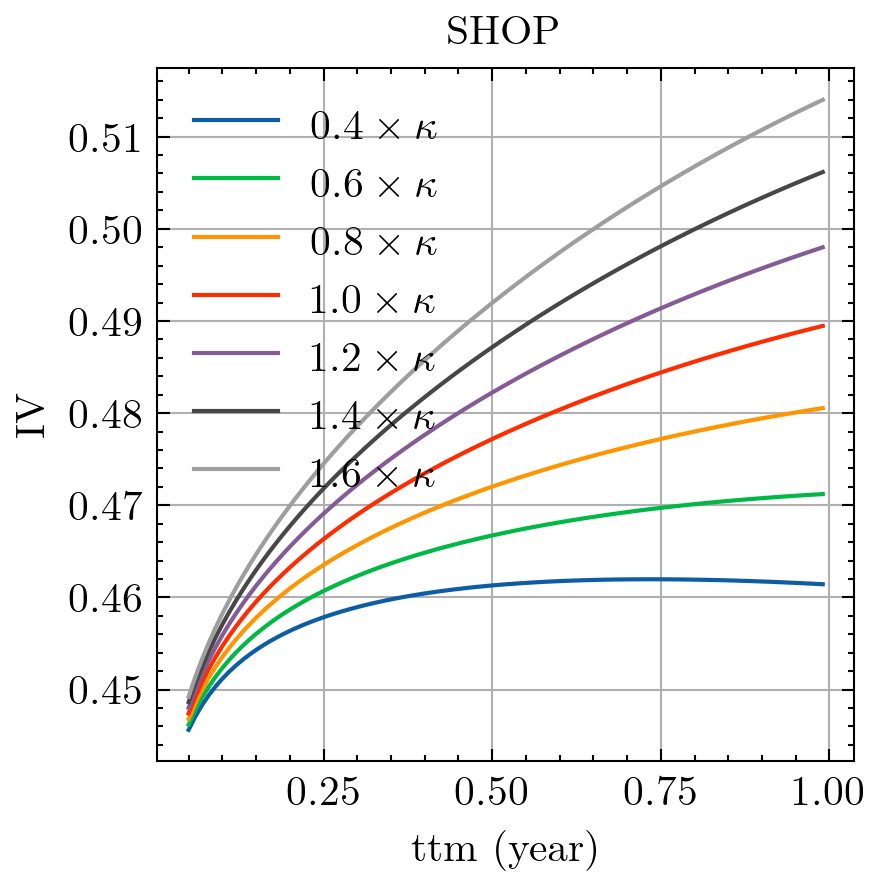}
      \caption{\footnotesize{Bumping $\theta$ - Effects on the term structure.}  }  \end{subfigure}\\
    \caption{\footnotesize{Effects of variations in $\kappa$ and $\theta$ on the term structure of the ATM implied volatility.}}
    \label{fig:term_structure_kappa_theta}
\end{figure}}

\subsubsection{Jump parameters}

{\begin{figure}
    \centering
    \begin{subfigure}{0.4\textwidth}
      \centering
      \includegraphics[width=\linewidth]{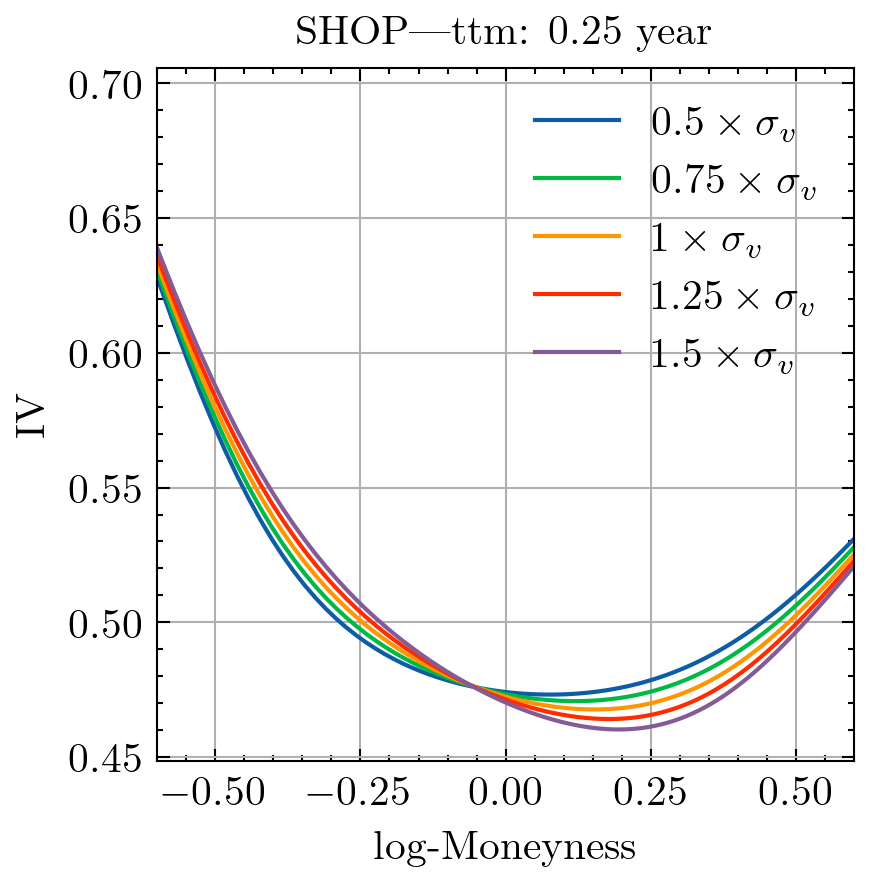}
      \caption{\footnotesize{Bumping $\sigma_v$ - maturity 3 months.}}
    \end{subfigure}
    \begin{subfigure}{0.4\textwidth}
      \centering
      \includegraphics[width=\linewidth]{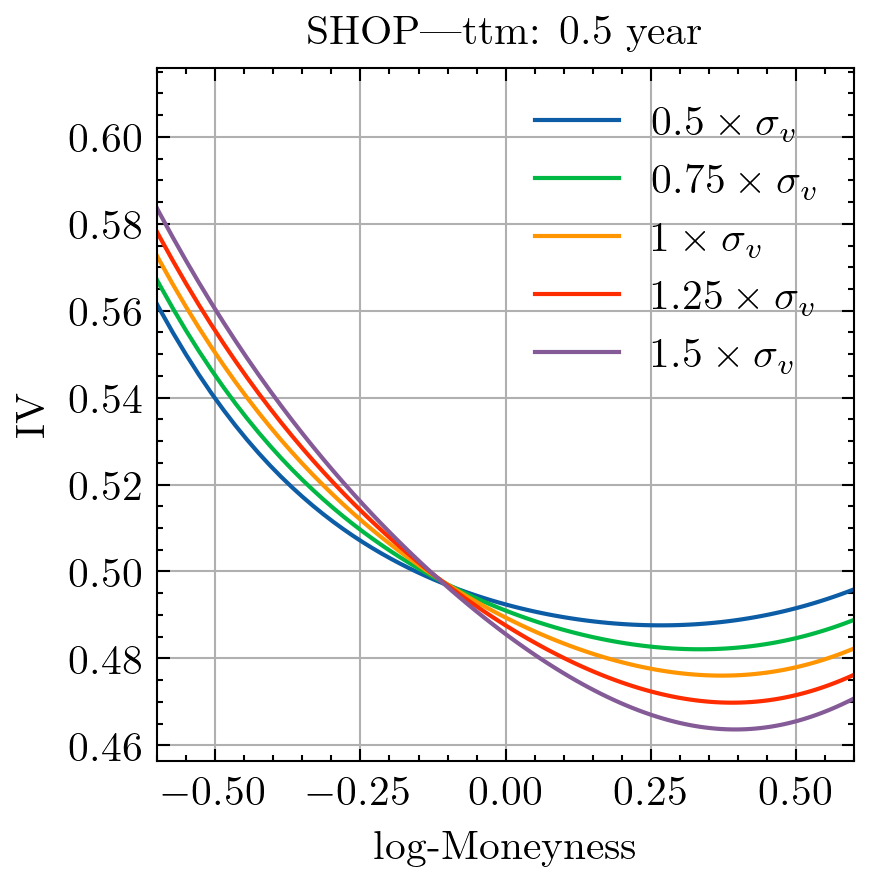}
      \caption{\footnotesize{Bumping $\sigma_v$ - maturity 6 months.}}   \end{subfigure}\\
    \caption{\footnotesize{Effects of variations in the parameter $\sigma_v$ on the implied volatility smiles. The plots represent implied volatility curves, for maturities of 3 and 6 months.}}
    \label{fig:bumping_sigma_v}
\end{figure}}

{\begin{figure}
    \centering
    \begin{subfigure}{0.4\textwidth}
      \centering
      \includegraphics[width=\linewidth]{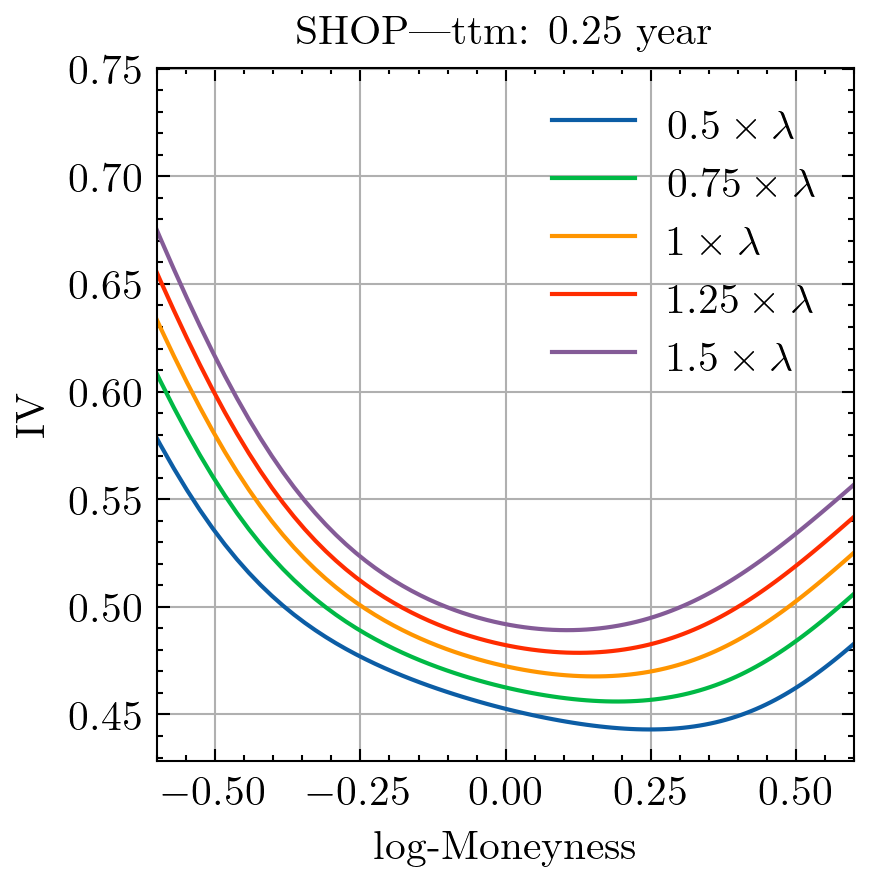}
      \caption{\footnotesize{Bumping $\lambda$ - maturity 3 months.}}
    \end{subfigure}
    \begin{subfigure}{0.4\textwidth}
      \centering
      \includegraphics[width=\linewidth]{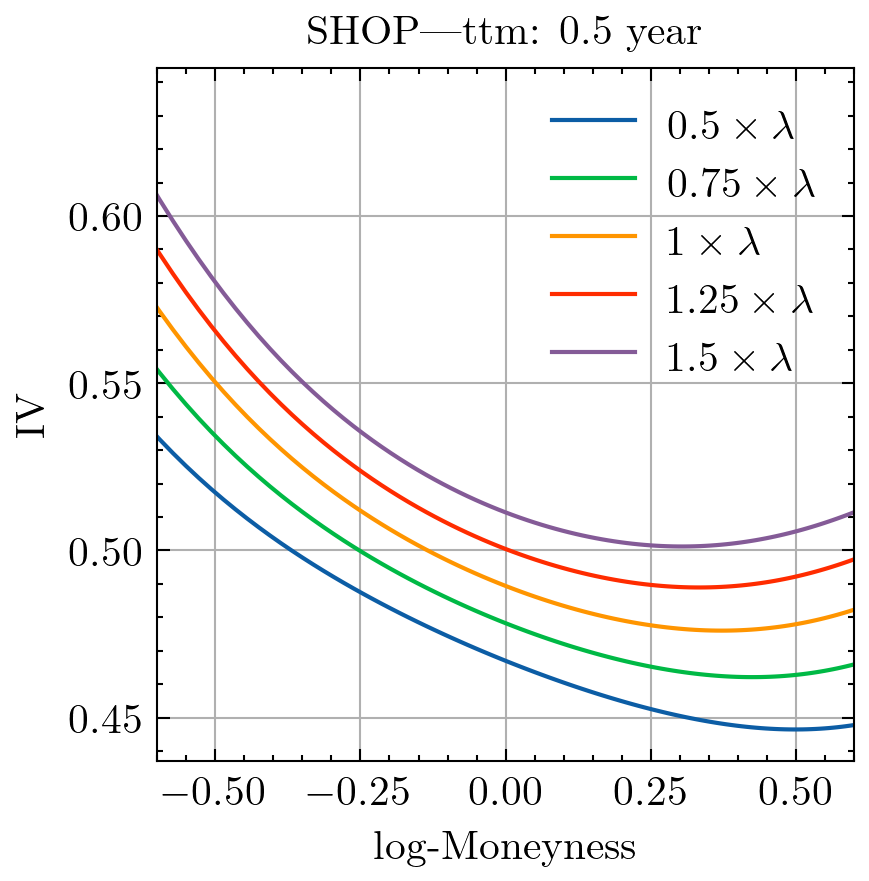}
      \caption{\footnotesize{Bumping $\lambda$ - maturity 6 months.}}   \end{subfigure}\\
    \caption{\footnotesize{Effects of variations in the parameter $\lambda$ on the implied volatility smiles. The plots represent implied volatility curves, for maturities of 3 and 6 months.} }
    \label{fig:bumping_lambda}
\end{figure}}

{
\begin{figure}
    \centering
    \begin{subfigure}{0.4\textwidth}
      \centering
      \includegraphics[width=\linewidth]{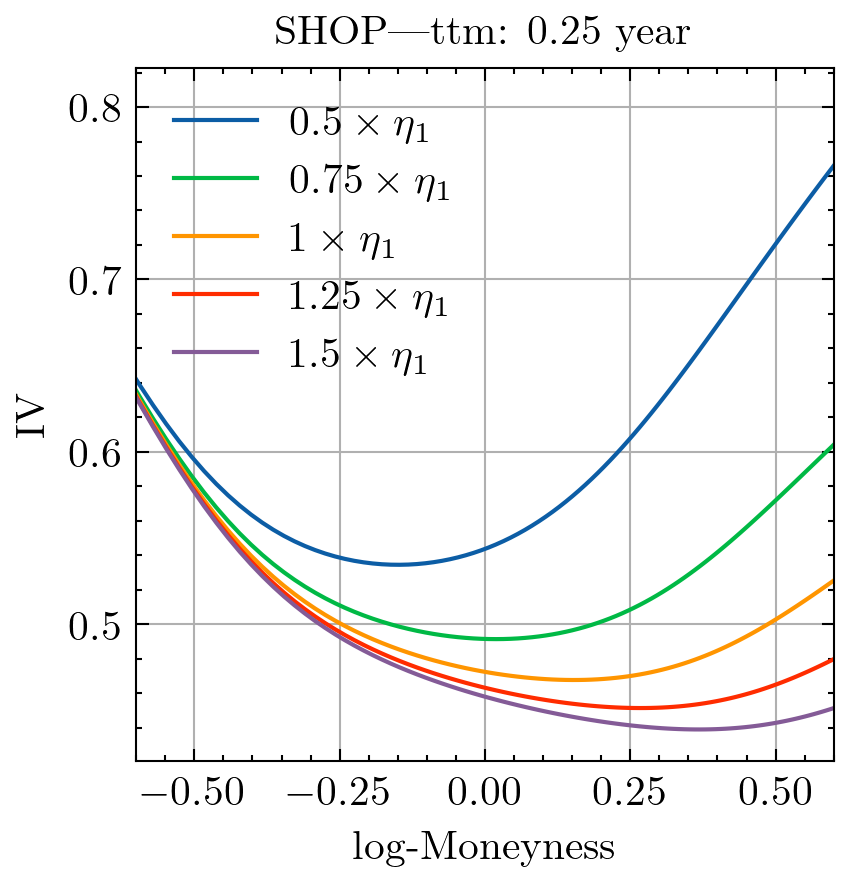}
      \caption{\footnotesize{Bumping $\eta_1$ - maturity 3 months.}}
    \end{subfigure}
    \begin{subfigure}{0.4\textwidth}
      \centering
      \includegraphics[width=\linewidth]{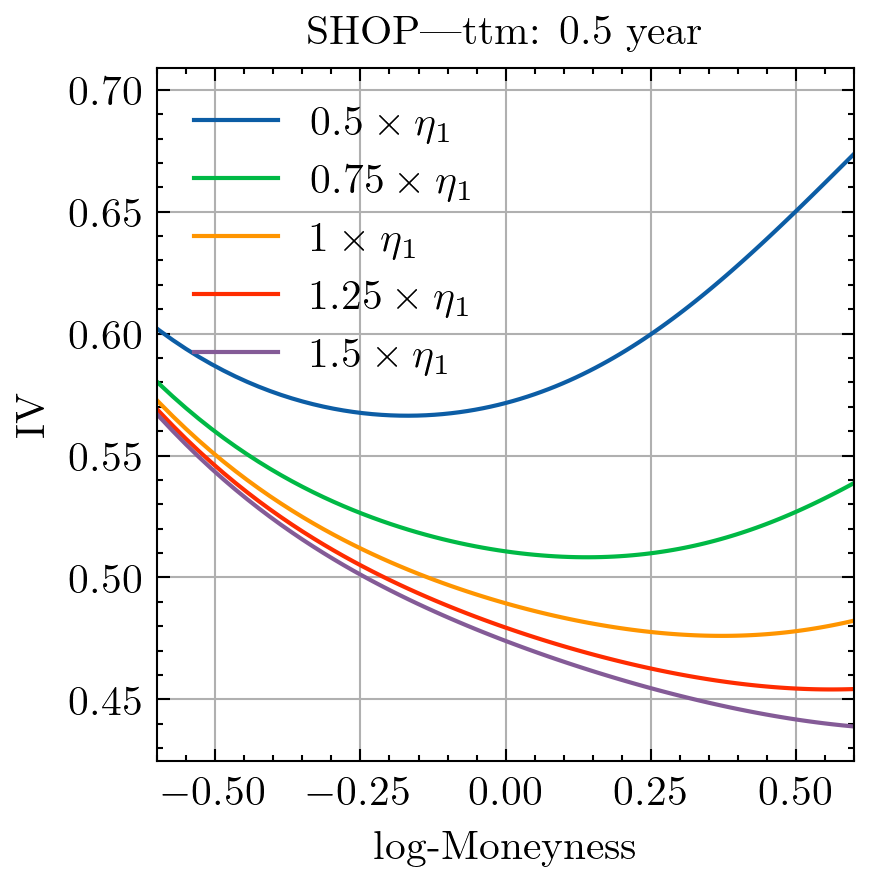}
      \caption{\footnotesize{Bumping $\eta_1$ - maturity 6 months.}}
    \end{subfigure}\vspace{0.5cm}\\
    \begin{subfigure}{0.4\textwidth}
      \centering
      \includegraphics[width=\linewidth]{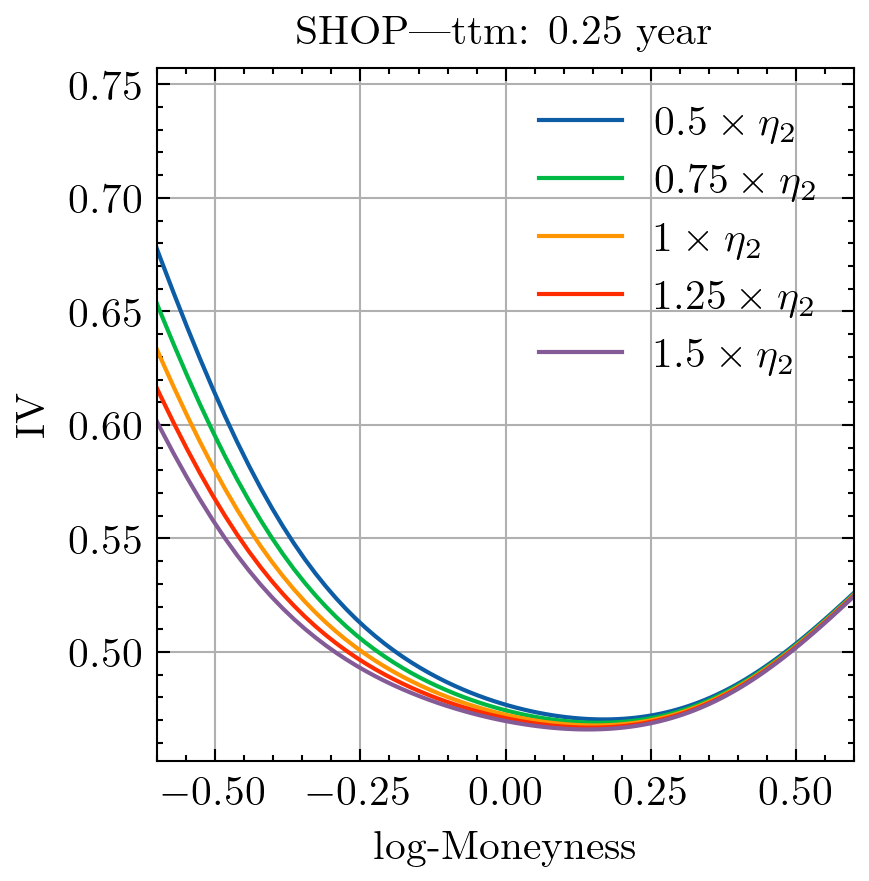}
      \caption{\footnotesize{Bumping $\eta_2$ - maturity 
    3 months.}}
    \end{subfigure}
    \begin{subfigure}{0.4\textwidth}
      \centering
      \includegraphics[width=\linewidth]{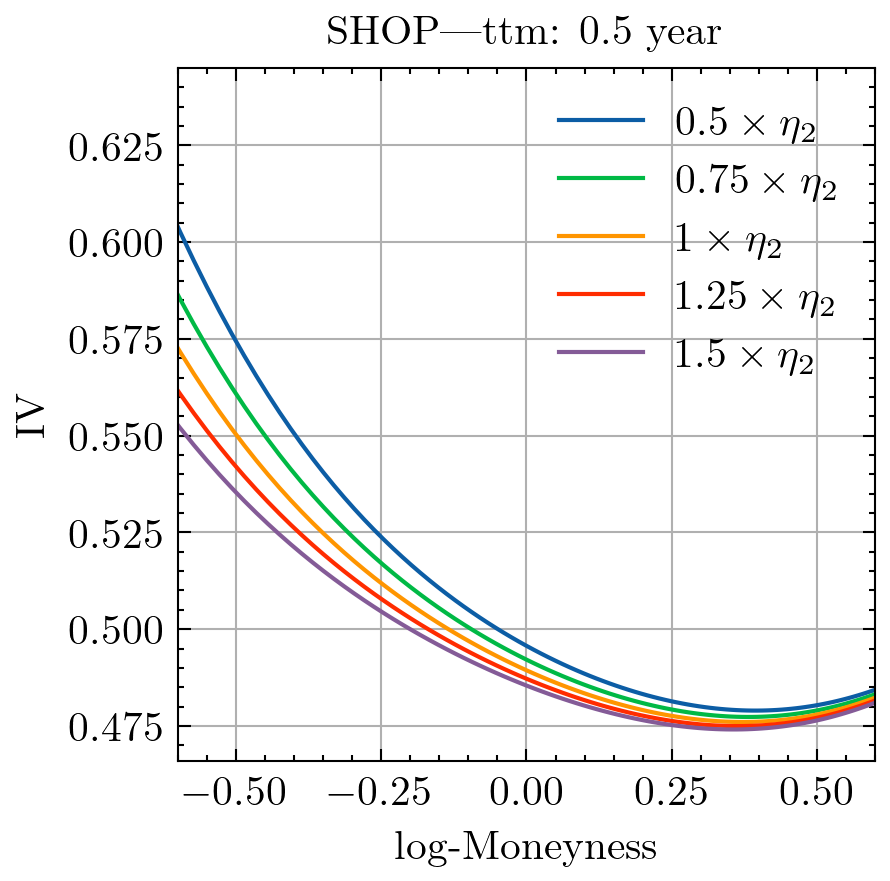}
        \caption{\footnotesize{Bumping $\eta_2$ - maturity 6 months.}}   \end{subfigure}\\
    \caption{
    \footnotesize{Effects of variations in the parameters $\eta_1$ and $\eta_2$ on the implied volatility smiles. The plots represent implied volatility curves, for maturities of 3 and 6 months. }
    }
\label{fig:bumping_etas}
\end{figure}
}

First, we note (unsurprisingly) that increasing the jump intensity $\lambda$, which controls the frequency of observed jumps, shifts the smile upward for all levels of moneyness (see Figure \ref{fig:bumping_lambda}). Second and more interestingly, the jump parameters $\eta_1$ and $\eta_2$ have a more significant and subtle impact on the shape of the implied volatility smile. Recall that the smaller the parameter $\eta_1$, the greater the magnitude of positive jumps; this increase in the jump size triggers a significant increase in the implied volatility for near ATM and OTM call options (see Figure \ref{fig:bumping_etas}). The same effect holds for negative jumps driven by the parameter $\eta_2$. However, because the calibrated value of $p$ from Table \ref{table:CalibParams} implies a much higher probability of positive jumps ($p=0.958$), the impact of negative jumps is more moderate compared to positive ones. It is therefore clear that the presence of two distinct tail heaviness parameters allows the HKDE model to capture more information from deep ITM and OTM prices, by adapting each wing of the implied volatility smile independently. This constitutes a clear improvement compared to the Bates model, where the symmetric distribution of jump size (log-normal) does not allow for a proper generation of asymmetric wings. As the wings are typically the hardest feature of the smile for a model to match, this places additional pressure on the other parameters of the Bates model to match the market smiles.

\section{Option Surface Pricing and Calibration}\label{sect:SurfacePriceCali}
This section briefly summarizes our calibration approach for HKDE and the challenger models. We start by describing the pricing approach in Section \ref{sect:biorproj} next, followed by the calibration methodology in Section \ref{sect:CalibMethod}. All source-code for this section is made publicly available as referenced below.

\subsection{Pricing via PROJ method}\label{sect:biorproj}

 {\color{black} 

Fourier pricing \cite{CarrMadanFourier} and its multiple refinements such as the COS \cite{CosineFang}, or PROJ \cite{kirkby2015efficient} methods are particularly well suited to the HKDE model, given the closed form of the HKDE characteristic function \eqref{Phi_HKDE}. We choose to implement the PROJ method, given its efficient and accurate performance even for late generation of exotics.

Let us briefly recall the PROJ method and the associated pricing formula for European style claims. The aim of PROJ method is to recover the density by projecting it into a B-splines space. The coefficients are obtained thanks to the isometric property of the Fourier transform. More precisely, let us consider a random variable $X$ and its known characteristic function $\phi$. We define a \textit{generator} $\varphi$ and a \textit{resolution} $a>0$ that determines the quality of the approximation. The function base $\varphi_{a,n}(x)$ is then the sequence of shifted generators, \textit{i.e.} for a uniform grid $\{x_n\}$, it is defined as:
\begin{equation}
    \varphi_{a,n}(x) := a^{1/2}\varphi(a(x-x_n)).
\end{equation}
If the generator is a B-spline function, the generated basis $(\varphi_{a,n})_{n\in\mathbb{Z}}$ is \textit{non-orthogonal} because composed of positive overlapping functions. It follows that the orthogonal projection $\bm{\pi}_a^\perp f$ of the density of $X$ onto $\overline{\mathrm{span}}\left[(\varphi_{a,n})_{n\in\mathbb{Z}}\right]$ is given by:
\EQ
    \label{eq:projection_span}
    \bm{\pi}_a^\perp f =  \sum_{n\in\mathbb{Z}} \langle f, \widetilde{\varphi}_{a,n}\rangle\varphi_{a,n}
\EN
where $(\widetilde{\varphi}_{a,n})_{n\in\mathbb{Z}}$ is the dual basis of $(\varphi_{a,n})_{n\in\mathbb{Z}}$. It is worth noting that $(\widetilde{\varphi}_{a,n})_{n\in\mathbb{Z}}$ and $(\varphi_{a,n})_{n\in\mathbb{Z}}$ are \textit{bi-orthogonal}, \textit{i.e.} $\langle\widetilde{\varphi}_{a,n} , \varphi_{a,m}\rangle = \delta_{n,m}$, and that, by the asymptotic density of $(\varphi_{a,n})_{n\in\mathbb{Z}}$, $||\bm{\pi}_a^\perp f-f||\to_{a\to 0} 0$. The projection coefficients $\langle f, \widetilde{\varphi}_{a,n}\rangle$ are available in closed form:
\EQ
\label{eq:density-proj-coef}
\langle f, \widetilde{\varphi}_{a,n}\rangle := \beta_{a,n} = \frac{1}{\pi\sqrt{a}} \mathrm{Re} \left[\int_0^\infty e^{-\mi x_n\xi}\phi(\xi)\widehat{\widetilde{\varphi}}_{a,n}\left(\frac{\xi}{a}\right)\D \xi \right]
\EN
and only require the explicit form of $\phi$ and $\widehat{\widetilde{\varphi}}$.

Given the definition of $\bm{\pi}_a^\perp f$, if a claim has a payoff of the form $g\left(S_T\right)$, its value $V_T$ can be then approximated by using the projected density, so that:
\EQ\label{eq: proj_analytical}
V_T \approx e^{-rT}\sum_{n\in\mathbb{Z}}\langle f, \widetilde{\varphi}_{a,n}\rangle \int_{x_{n-1}}^{x_{n+1}}\varphi_{a,n}(y) g(S_0e^y)\D y.
\EN
since $\varphi_{a,n}$ is supported on $[x_{n-1},x_{n+1}]$.
\begin{remark}\label{rem:projection}
In numerical pricing, we set a leftmost boundary grid point $x_1$, and a gridwidth of $2\bar \alpha$ is chosen to encompass the support of $f$. A number $N$ of basis elements is then selected, which controls the approximation accuracy, and restricts the span of the basis elements to the finite set $\{\varphi_{a,n}\}_{n=1}^N$, covering $[x_1,x_1+2\bar \alpha]$. 

Each $\varphi_{a,n}(x)$ is centered over the grid point $x_n = x_1 +(n-1)\Delta_a$, where $\Delta_a :=1/a = 2\bar\alpha / (N-1)$.  
Similar to \cite{fang2009novel}, $\bar\alpha$ is selected using the cumulant-based rule,
\begin{equation}\label{eq:Alphabar}
\bar\alpha \equiv \bar\alpha(L_1, t):=  \max\Big\{\frac{1}{2}, \text{ }L_1\cdot \sqrt{ c_2 t + \sqrt{c_4 t}}\Big\},
\end{equation}
where $c_i$ is the $i^{th}$ cumulant of the log return process for $t=1$, and $L_1 = 10\sim 30$ is fixed. 

The final approximation of the density is given by
\begin{equation}\label{eq:fbarPROJ}
f(x) \approx  a^{1/2}\Upsilon_{a,N}\sum_{1\leq k\leq N}\bar\beta_{a,k} \cdot \varphi_{a,k}(x),
\end{equation}
with the constant $\Upsilon_{a,N}$ depending on the order of the B-spline function (values are available in \cite{kirkby2015efficient}). Finally, \eqref{eq:fbarPROJ} is  used to approximate \eqref{eq:density-proj-coef}, and the coefficients $a^{1/2}\Upsilon_{a,N}\cdot\bar \beta_{a,k} \approx \beta_{a,k}$
 are computed through the exponentially convergent discretization. Further details about projection coefficients are reported in Appendix \ref{app:implementation}.
\end{remark}
{\color{black}
\subsection{Calibration Methodology}\label{sect:CalibMethod}

Our calibration methodology relies on the PROJ method for its speed, accuracy and robustness for short maturity options (see \cite{kirkby2015efficient}). Each model we will test is calibrated to fit market prices for several tenors. 
Let denote $(\mathcal{T}_n)_{(1\leq n \leq N)}$ the set of available maturities, $(N_n)_{(1\leq n \leq N)}$ the number of options for each of these tenors, $(v_j^{(n)})_{1\leq j \leq N_n} $ the option prices with strikes $(K_j^{(n)})_{1\leq j \leq N_n} $ and $(\sigma_j^{(n)})_{1\leq j \leq N_n} $  the corresponding implied volatilities. We will calibrate the models only on OTM options by convention. The calibration consists of a classical least squares problem:
\EQ 
    \bm{P}^* \coloneqq \arg\min_{P}\left[ \sum_{n=1}^N\sum_{j=1}^{N_n}w_j^{(n)}\left( \mathcal{V}_j^{(n)}(P)-v_j^{(n)}\right)^2\right],
\EN
where $\mathcal{V}_j^{(n)}(P)$ denotes the model price of an option with maturity $\mathcal{T}_n$ and strike $K_j^{(n)}$, under the set of parameters $P$. We use the well-known \textit{vega-weighting} (see for example \cite{vegaWeightHeston}):
\EQ
    w_j^{(n)} := \left(S_0\psi(d_1)\sqrt{\mathcal{T}_n}\right)^{-1}, \quad d_1 := \frac{1}{\sigma_j^{(n)}\sqrt{\mathcal{T}_n}}\left(\log \frac{S_0}{K_j^{(n)}}+\mathcal{T}_n\left(r_n-q_n+\frac{1}{2}\left(\sigma_j^{(n)}\right)^2\right)\right),
\EN
where $r_n$ and $q_n$ denote the risk-free rate and dividend yield at maturity $\mathcal{T}_n$. \\
$\psi$ represents the probability density function of a standard normal distribution. The minimization problem is numerically solved using the {\bf\sffamily fypy} repository\footnote{\url{https://github.com/jkirkby3/fypy}}. Specifically, for each stochastic model subjected to minimization, an iterative calibration process is employed, gradually increasing the precision required for the stopping criterion in the minimization routine. \\
The minimization process for each iteration is executed using the Trust Region Reflective algorithm, as implemented in \texttt{scipy.optimize.least\_squares}. 
The stopping criterion is defined as the tolerance level for termination based on changes in the cost function.
The initial guess used for each iteration is the set of the calibrated parameters obtained in the preceding iteration.\\
Using the above-mentioned algorithm within the iterative procedure ensures a balance between computational efficiency and accuracy at each stage of the calibration. The iterative refinement process benefits from the algorithm’s stability, ability to handle parameter constraints effectively, and capacity to achieve high precision in cost function minimization. The overall iterative structure allows to incorporate improvements progressively.

\section{Market Calibration Results}\label{sec:MarketCalibrationResults}
This section provides several experiments demonstrating the quality of fit of the HKDE model. 
The experimental setup consists of four well-known single name equity assets, namely Amazon (AMZN), Netflix (NFLX), Shopify (SHOP), and Spotify (SPOT). Each of these assets is characterized by steep implied volatility smiles, which present a challenging setup for model calibration. The data used for calibration was extracted on the 20-th of February 2024 and the calibration methodology has been documented in Section \ref{sect:CalibMethod}. 

To analyze the model in the proper context, we compare it with the natural benchmark of Heston, which has no jump component, as well as Bates, which is perhaps the most widely used SVJ model and shares in common the Cox-Ingersol-Ross (CIR) model for the stochastic volatility component that underpins Heston.  While the focus of this work is stochastic volatility, we also choose a L\'evy model for comparison - the Bilateral Gamma Motion (BGM), see for example \cite{aguilarkirkby2022robust}. This choice is based on recent findings showing this model's consistent calibration outperformance compared to other L\'evy models, such as NIG and Variance Gamma, as well as common jump-diffusion models, while demonstrating similar performance to CGMY. See \cite{kirkAglAgu24} for a recent analysis.  

{
\begin{figure}
    \centering
    \begin{subfigure}{0.43\textwidth}
      \centering
      \includegraphics[width=\linewidth]{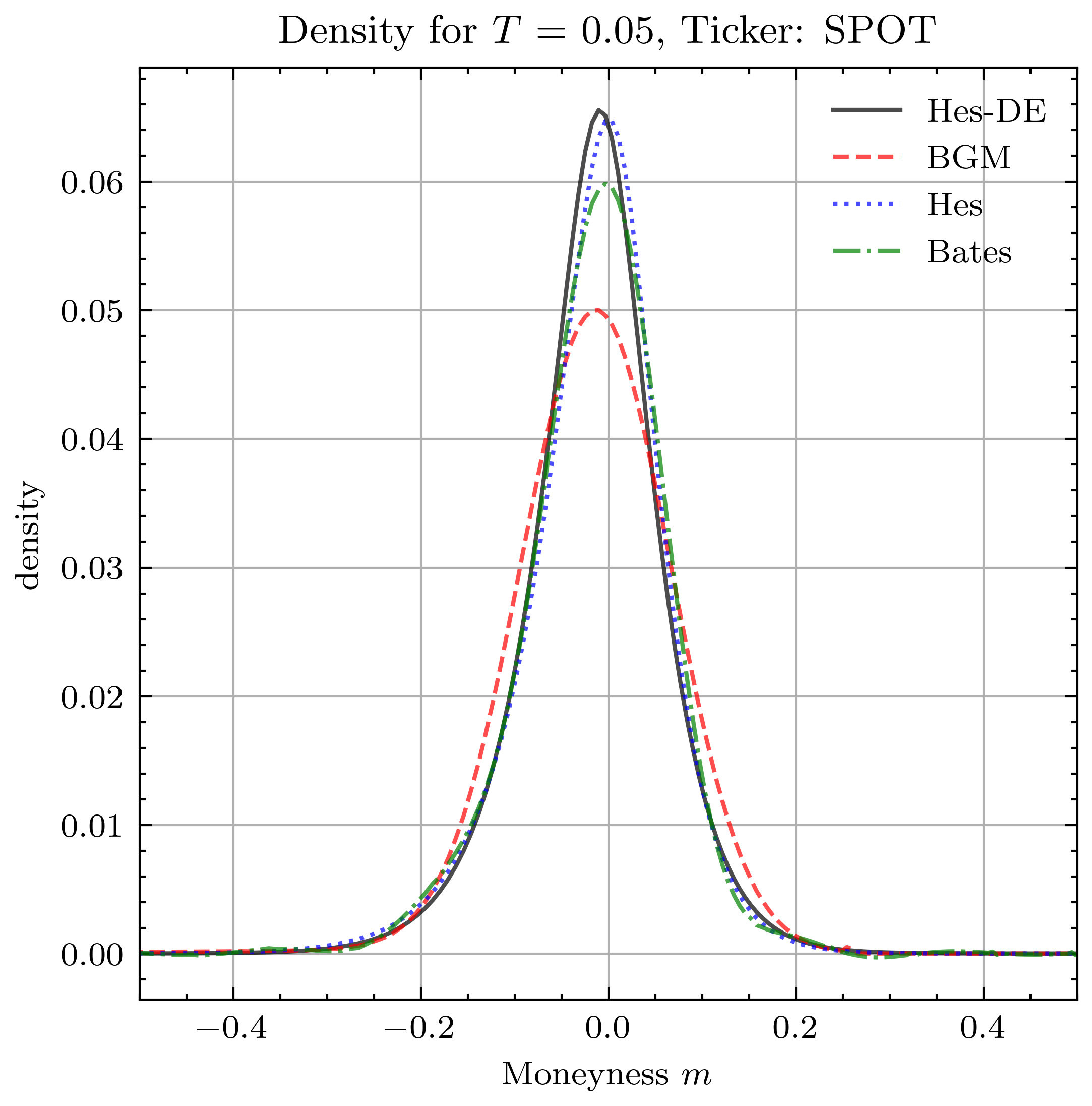}
      \caption{\footnotesize{$T=0.05$}}
    \end{subfigure}
    \begin{subfigure}{0.43\textwidth}
      \centering
      \includegraphics[width=\linewidth]{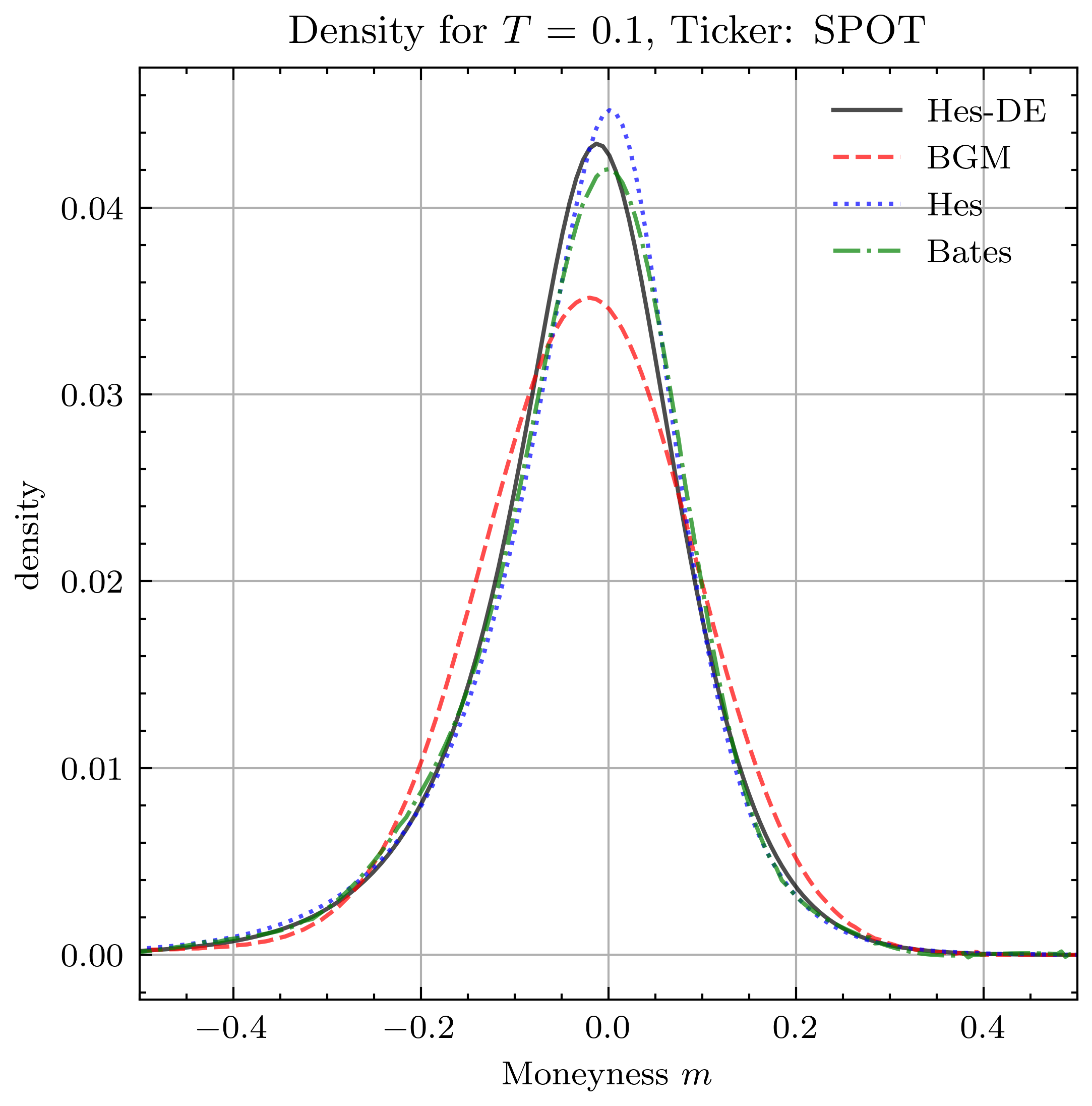}
      \caption{\footnotesize{$T=0.1$}}
    \end{subfigure}\\ \vspace{0.5cm}
    \begin{subfigure}{0.43\textwidth}
      \centering
      \includegraphics[width=\linewidth]{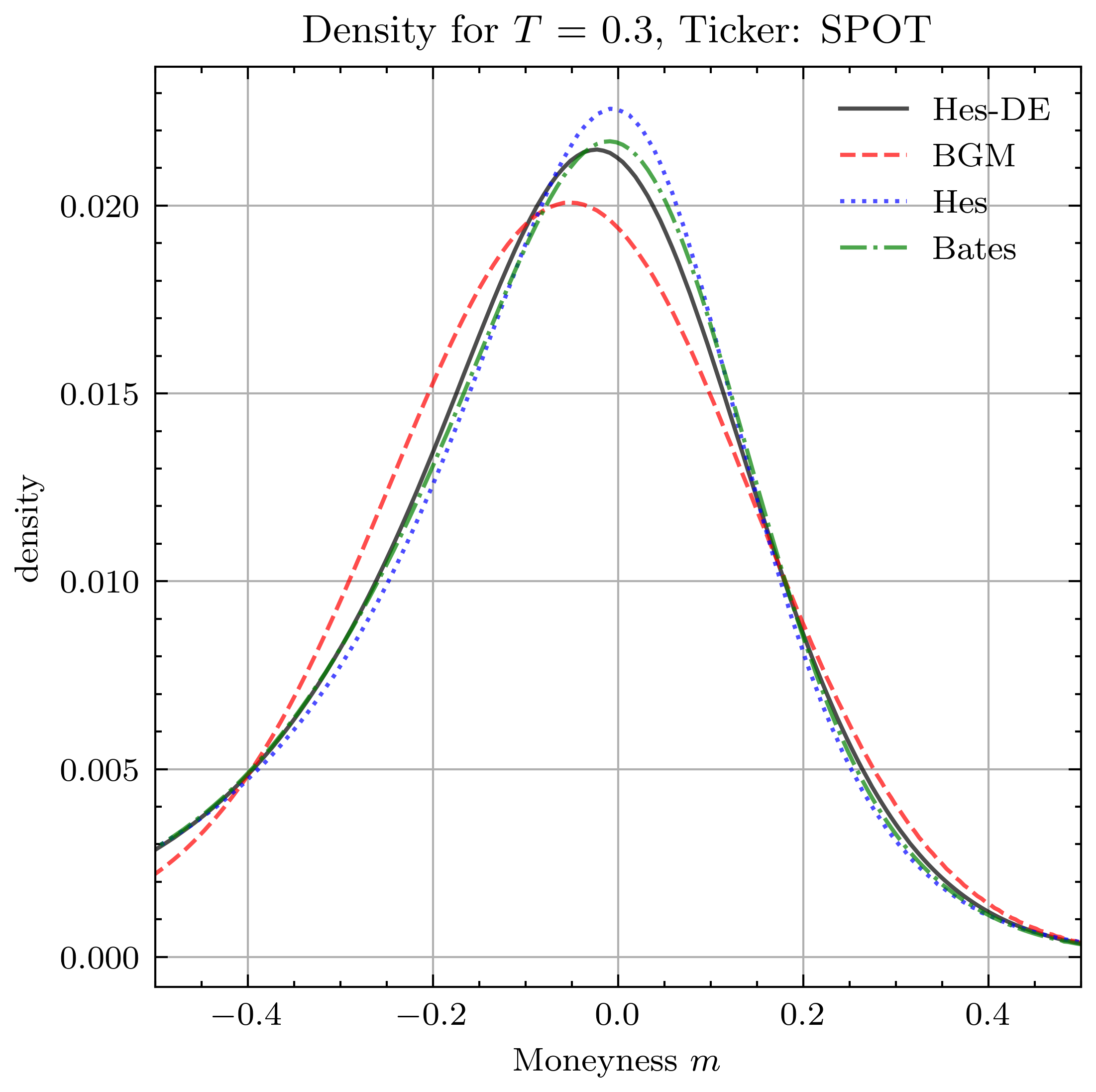}
      \caption{\footnotesize{$T=0.3$}}
    \end{subfigure}
    \begin{subfigure}{0.43\textwidth}
      \centering
      \includegraphics[width=\linewidth]{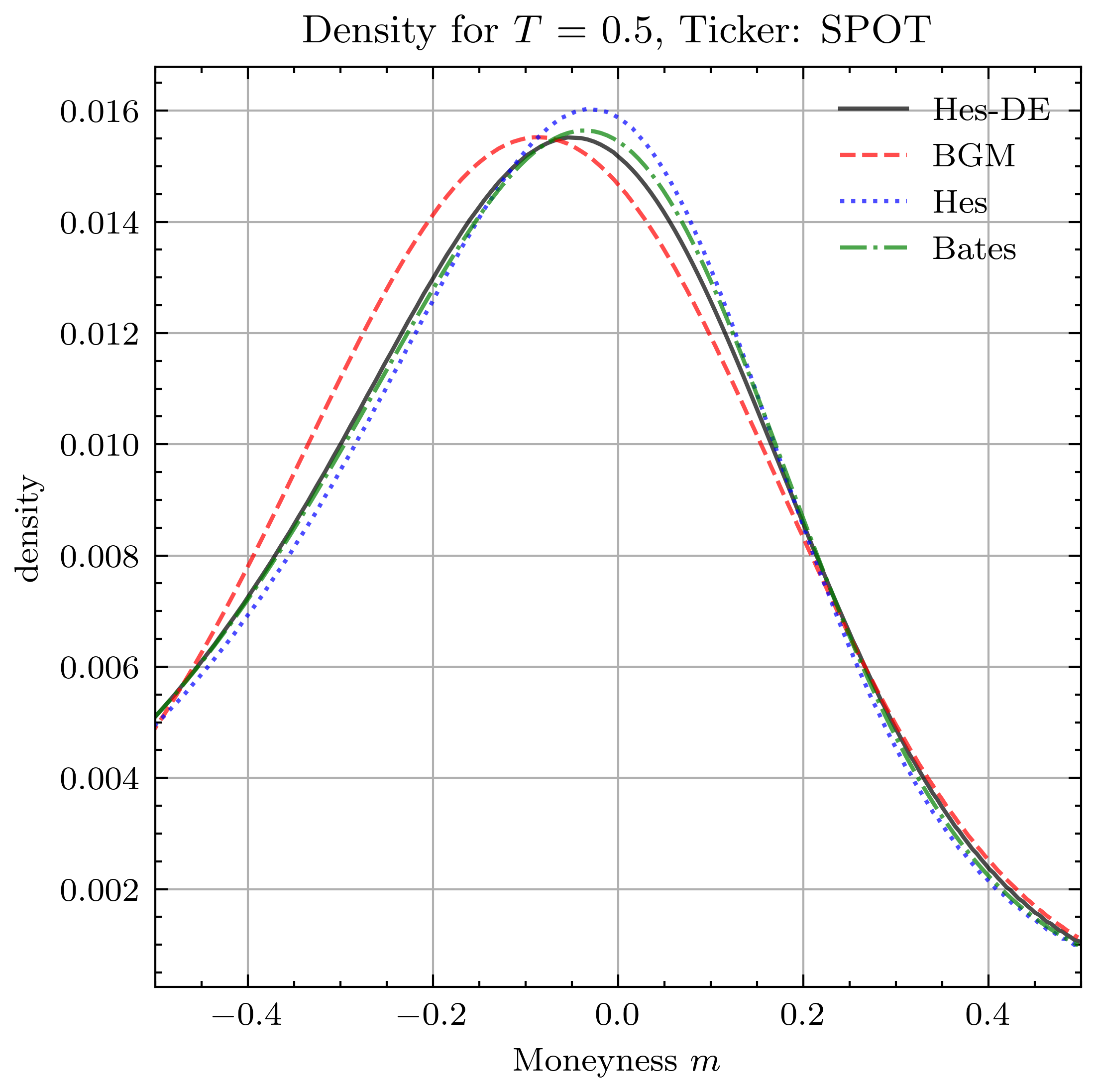}
      \caption{\footnotesize{$T= 0.5$}}
    \end{subfigure}\\
    \caption{\footnotesize{
    Transition densities of the calibrated models for the ticker SPOT and maturities $T\in\{0.05,0.1,0.3,0.5\}$.}
    }
    \label{fig:transition_dens}
\end{figure}
}

{\footnotesize
\begin{table}[h!t!b!]
    \centering 
    \scalebox{1}{
    \begin{tabular}{|c|c|lllllllll|}
    \hline
     \textbf{Model} & \textbf{Underlying} & \multicolumn{9}{c|}{\textbf{Parameters}}\\ \hline
    
    

    BGM & -- &   $\alpha_+$ & $\lambda_+$ & $\alpha_-$ &$\lambda_-$  & $\sigma$ & &   & &  \\  \hline
     & AMZN  &  3.093 & 22.88 & 0.415 & 3.342 & 0.248& &&    & \\
     & NFLX  & 0.032 & 258.818 & 0.184 & 2.017 & 0.316 &    &&   &   \\
     & SHOP  & 6.165 & 11.075 & 3.201 & 4.34 & 0.265 &  &&      &    \\
     & SPOT & 14.706 & 238.363 & 0.168 & 1.839 & 0.351 &  &&  &  \\
      \hline
    
    HKDE & -- &   $V_0$ & $\theta$ & $\kappa$ &$\sigma_v$  &  $\rho$&$\lambda$ &$p$   &$\eta_1$ & $\eta_2$  \\  \hline
      & AMZN  & 0.023 & 0.067 & 5.275 & 1.268 & -0.691 & 53.165 & 0.999 & 49.799 & 2.587 \\
     & NFLX  & 0.001 & 0.091 & 13.355 & 4.797 & -0.498 & 103.622 & 0.272 & 42.945 & 65.011  \\
     & SHOP & 0.176 & 0.728 & 0.191 & 0.194 & -0.718 & 1.009 & 0.958 & 8.739 & 0.733   \\
     & SPOT &  0.064 & 0.163 & 6.796 & 1.698 & -0.391 & 17.725 & 1.0 & 35.555 & 0.049 \\
      \hline
    
    Heston & -- &   $V_0$ & $\theta$ & $\kappa$ &$\sigma_v$  & $\rho$ & &   & &  \\  \hline
     & AMZN  & 0.062 & 0.109 & 14.825 & 3.077 & -0.264 & &&    & \\
     & NFLX  & 0.066 & 0.151 & 14.857 & 2.987 & -0.279 &    &&   & \\
     & SHOP  & 0.216 & 0.268 & 43.472 & 10.0 & -0.183 &  &&      &\\
     & SPOT & 0.094 & 0.199 & 6.95 & 2.133 & -0.23 &  &&  & \\
      \hline
    
    Bates & -- &   $V_0$ & $\theta$ & $\kappa$ &$\sigma_v$  &  $\rho$&$\lambda$ &$\mu_J$   &$\sigma_J$ &   \\  \hline
     & AMZN  &  0.07 & 0.113 & 3.46 & 0.809 & -0.299 & 0.021 & -0.37 & 0.635&   \\
     & NFLX  &  0.067 & 0.146 & 14.254 & 2.434 & -0.275 & 0.002 & -9.343 & 3.901&  \\
     & SHOP  & 0.192 & 0.221 & 49.841 & 5.093 & -0.075 & 0.051 & -1.014 & 1.073&    \\
     & SPOT & 0.094 & 0.191 & 6.344 & 1.617 & -0.258 & 0.002 & -40.123 & 8.946&     \\
      \hline  
      
    \end{tabular}
    }
    
    \caption{\footnotesize{Calibrated parameters for the HKDE model and its challengers.}}
    \label{table:CalibParams}
\end{table}}}

\begin{table}[ht]
\centering
\footnotesize{
\begin{tabular}{|p{1cm}||p{3.5cm}|p{3.5cm}|p{3.5cm}|p{3.5cm}|}
\hline
\multicolumn{5}{|c|}{\textbf{Calibration Ranking - MAPE}} \\
\hline
\textbf{Rank} & \textbf{AMZN} & \textbf{NFLX} & \textbf{SHOP} & \textbf{SPOT} \\
\hline
1/4 & \textbf{HKDE (2.60974\%)} & Bates (3.56152\%) & \textbf{HKDE (2.65742\%)} & \textbf{HKDE (3.39366\%)} \\
\hline
2/4 & Bates (2.84991\%) & \textbf{HKDE (4.87563\%)} & Bates (4.01081\%) & Bates (3.73386\%) \\
\hline
3/4 & Heston (4.35531\%) & Heston (6.42533\%) & BGM (5.30184\%) & Heston (6.19707\%) \\
\hline
4/4 & BGM (4.92432\%) & BGM (7.75478\%) & Heston (5.32895\%) & BGM (8.93164\%) \\
\hline
\end{tabular}

\vspace{0.5cm}

\begin{tabular}{|p{1cm}||p{3.5cm}|p{3.5cm}|p{3.5cm}|p{3.5cm}|}
\hline
\multicolumn{5}{|c|}{\textbf{Calibration Ranking - RMSE}} \\
\hline
\textbf{Rank} & \textbf{AMZN} & \textbf{NFLX} & \textbf{SHOP} & \textbf{SPOT} \\
\hline
1/4 & \textbf{HKDE (0.01433)} & Bates (0.02975) & \textbf{HKDE (0.02938)} & \textbf{HKDE (0.03173)} \\
\hline
2/4 & Bates (0.01436) & BGM (0.05996) & Bates (0.03569) & Bates (0.03255) \\
\hline
3/4 & BGM (0.02808) & \textbf{HKDE (0.10048)} & BGM (0.05785) & BGM (0.06023) \\
\hline
4/4 & Heston (0.02987) & Heston (0.12327) & Heston (0.07532) & Heston (0.10503) \\
\hline
\end{tabular}
\caption{\footnotesize{Ranking of calibration MAPE and RMSE for HKDE models and its challengers.}}
\label{table:Ranking}
}
\end{table}

{\color{black}
\subsection{Calibrated Parameters} 
Table \ref{table:CalibParams} provides the calibrated parameters for the different models and tickers mentioned in the setup details. To illustrate the difference between calibrated models, Figure \ref{fig:transition_dens} displays the transition density of four selected tenors in the SHOP market. The three stochastic volatility models are fairly similar in overall shape, especially when compared with the BGM (L\'evy) model. However, the differences in tail behavior, which are hard to visualize in the transition density, are quite pronounced as will be seen later in the plots of implied volatility smiles.

Observing the differences between the calibrated parameters of the Heston, Bates, and HKDE models for the same underlying assets, several interesting considerations emerge.
Regarding the initial volatility $V_0$ and the long-term volatility $\theta$, the Heston and Bates models tend to show higher values compared to HKDE, with notable similarity between the first two, except for the SHOP underlying asset.
Among the most significant parameters to observe, $\sigma_v$, $\rho$, and $\lambda$, the HKDE model consistently provides lower values for $\sigma_v$, indicating lower volatility incorporated into the diffusion process. Similarly, HKDE regularly presents lower values for the correlation parameter and higher values for $\lambda$, placing greater emphasis on the jump component.

\subsection{Error Metrics} Table \ref{table:Ranking} displays the error metrics results, i.e. the Mean Absolute Percentage Error (MAPE) and the Root Mean Square Error (RMSE). The error metrics are calculated across the entire calibrated volatility surface, relative to the market implied volatilities.
It can be observed that the HKDE and Bates models are systematically the top performers and that, for three stocks (AMZN, SHOP and SPOT), the HKDE model is the best one in terms of both MAPE and RMSE.  In terms of RMSE, the Heston model performs the worst, and is  beaten out by the BGM model, although in general the performances of the BGM and Heston models are similar. Overall, the HKDE model delivers the best all-around calibration results, and is outperformed only in the NFLX market.\\
These results are quite promising and demonstrate that the HKDE model is able to compete with industry-standard models. Moreover, since the pricing and simulation methodologies for these two models are entirely similar, HKDE is an easily adoptable replacement or complement for the Bates model in practice.  

The improved calibration performance of the HKDE model is not surprising as it adds an additional calibratable parameter over the Bates model. However, the addition of such parameter is economically defensible on the grounds that it accounts for empirically observed differences in the magnitude of up and down shocks. Having an understandable (and stable) interpretation of the model parameters is important for adoption into practice, as risk managers must understand the role played by each parameter in the model, and how changes in those parameters should predictably impact pricing.  Having econometric interpretations also lends to estimation procedures based on historical data that can be used to support the parameters that are calibrated to the market surface, and can also provide reasonable bounds for their calibrated values. It is also not uncommon for risk managers to fix a subset of model parameters to their econometrically calibrated values in order to provide day-over-day stability in the model.
}

{
\begin{figure}
    \centering
    \begin{subfigure}{0.45\textwidth}
      \centering
      \includegraphics[width=\linewidth]{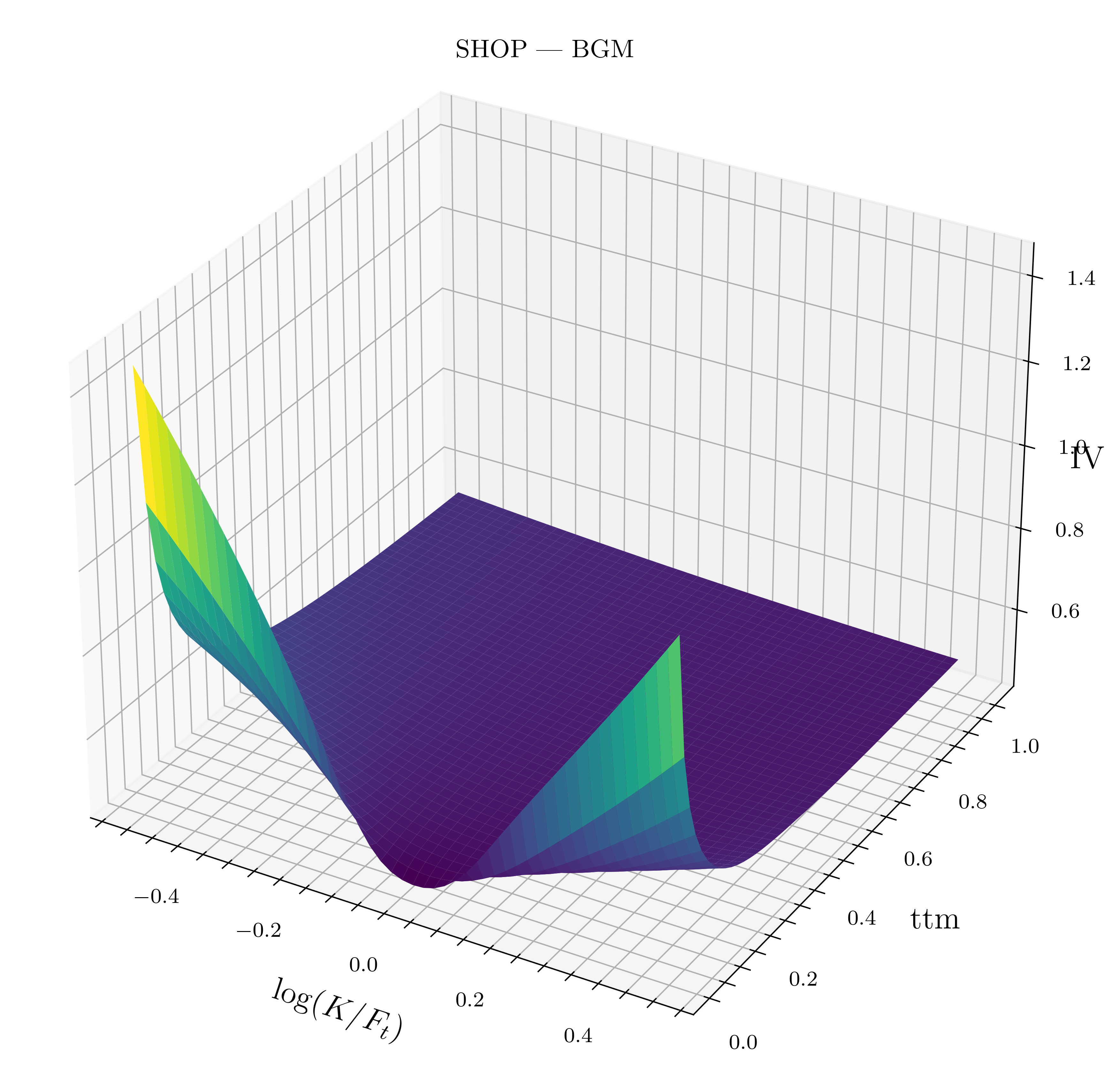}
      \caption{\footnotesize{BGM: SHOP}}
    \end{subfigure}
    \begin{subfigure}{0.45\textwidth}
      \centering
      \includegraphics[width=\linewidth]{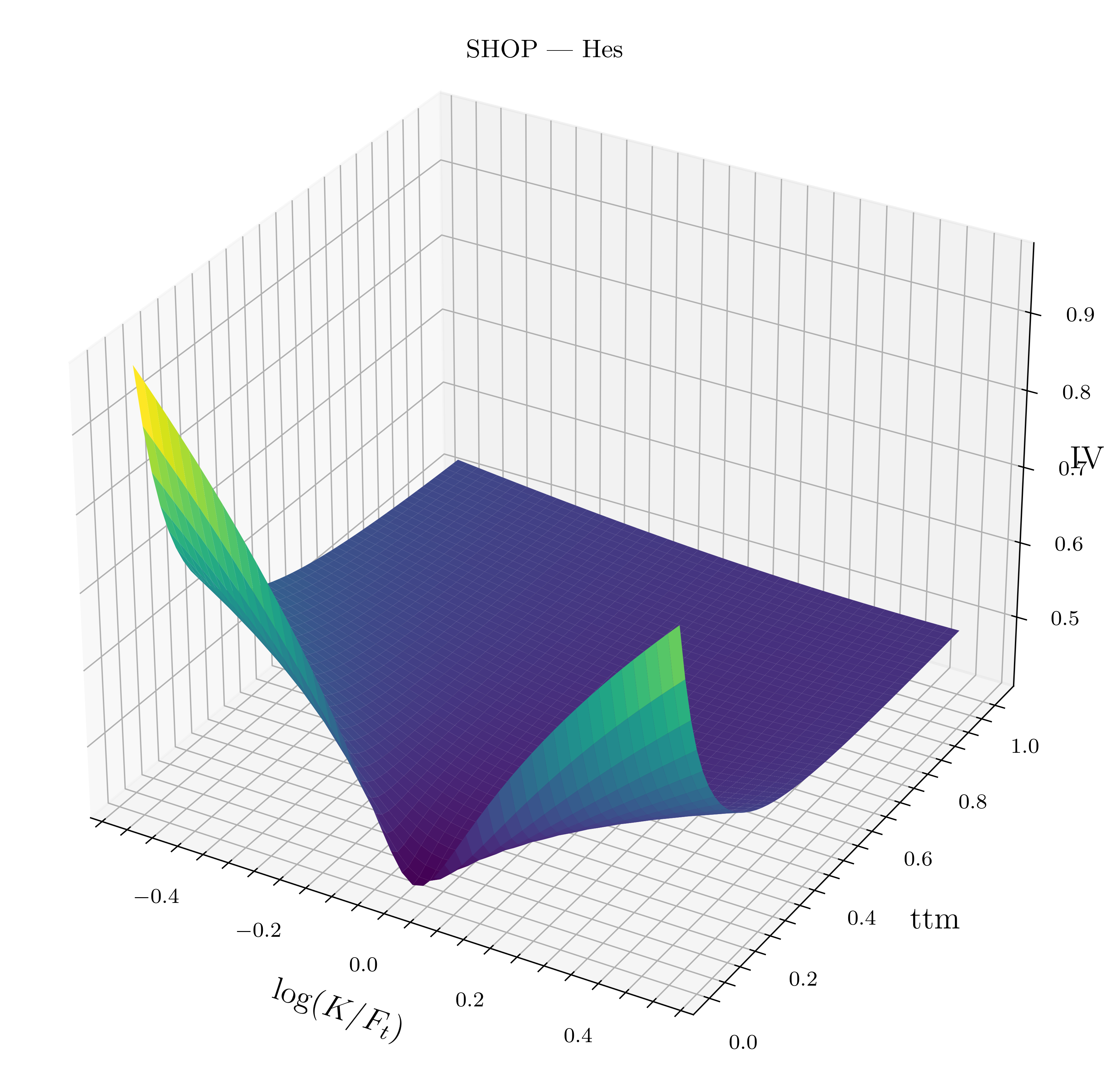}
      \caption{\footnotesize{Heston: SHOP}}
    \end{subfigure}\\
    \begin{subfigure}{0.45\textwidth}
      \centering
      \includegraphics[width=\linewidth]{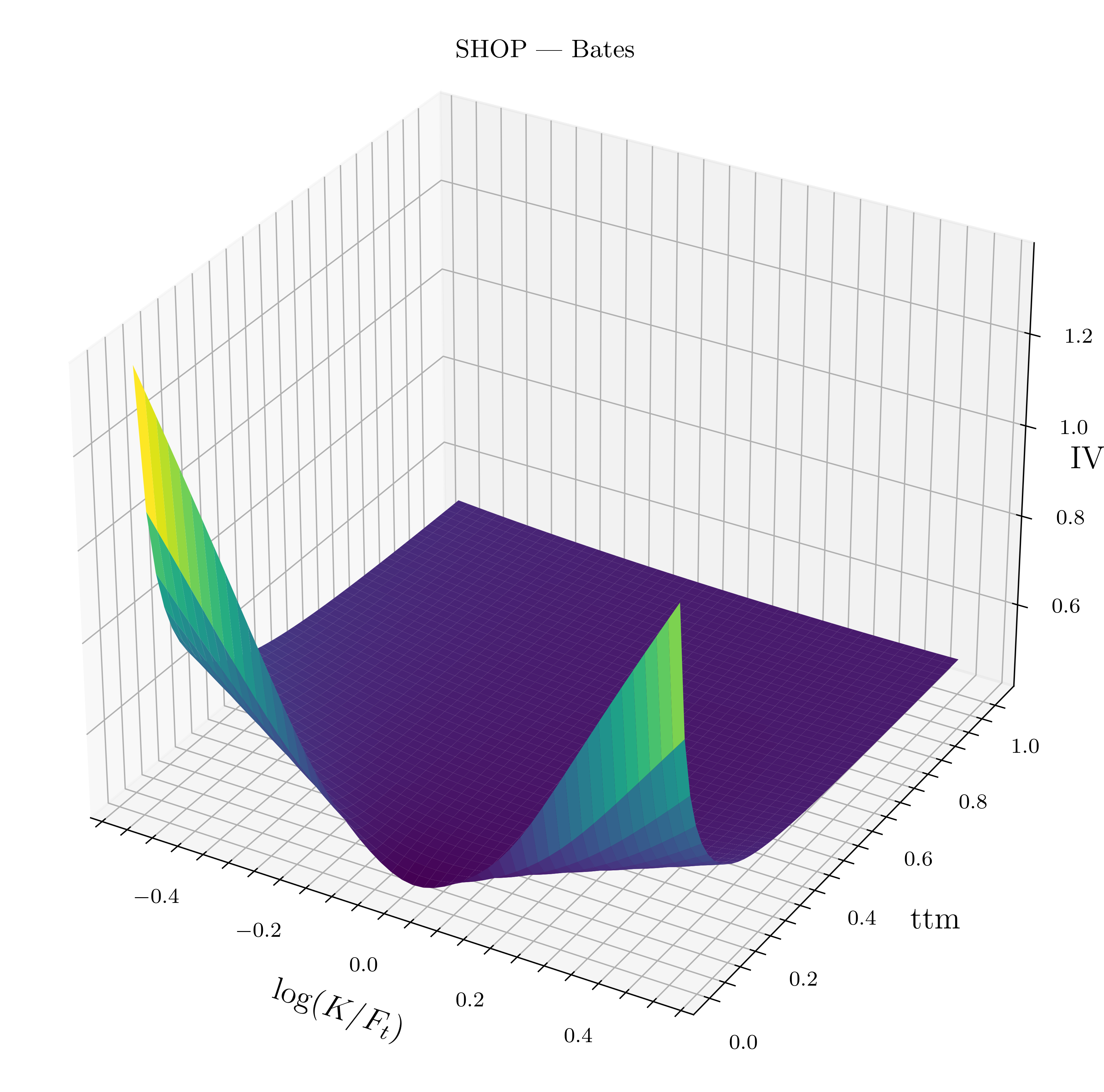}
      \caption{\footnotesize{Bates: SHOP}}
    \end{subfigure}
    \begin{subfigure}{0.45\textwidth}
      \centering
      \includegraphics[width=\linewidth]{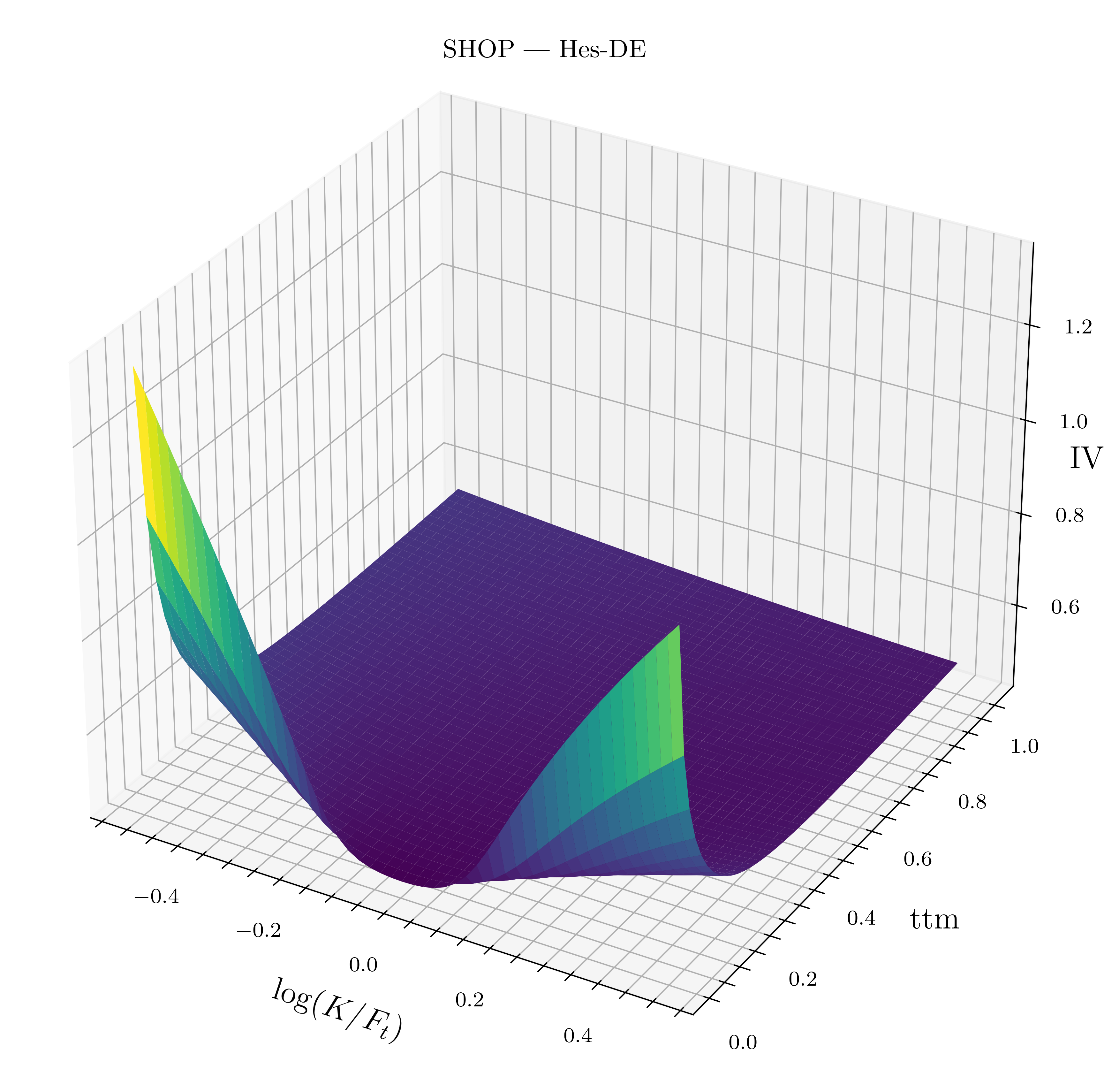}
      \caption{\footnotesize{Hes-DE: SHOP}}
    \end{subfigure}\\
    \caption{\footnotesize{Implied volatility surfaces under the four tested models for the underlying SHOP. }}
    \label{fig:3DIVSurface}
\end{figure}
}
\subsection{Visual Comparison with Challenger Models}
After analyzing the numerical results and demonstrating that the HKDE model exhibits the lowest calibration errors compared to its competitors, we next present a visual comparison of the models. Figure \ref{fig:3DIVSurface} illustrates the 3D plots of the implied volatility surfaces generated by the calibrated models for the SPOT market. From this perspective, the models appear quite similar, with steep near term smiles and a flattening of the surface volatility beyond a quarter to expiry. The real differences emerge upon inspection of the individual smiles, shown next.

We present in Figure \ref{fig:ImpliedVolSmile} the implied volatility smiles for the HKDE and competing models, which are cross-sections of the fitted implied volatility surfaces. 
This study examines four stocks (AMZN, NFLX, SPOT, SHOP) across three typical time-to-maturity (ttm) periods: 8 days, 57 days, and 148 days, representing short, mid, and long-term options.
There are several key features to observe. The superior performance of the HKDE model is particularly visible for short-term options, where the model fits the market smile almost perfectly, even in deep in-the-money (ITM) and out-of-the-money (OTM) regions. In comparison, in the negative moneyness region, the Heston and Bates models tend to have flatter wings that do not appropriately fit the market. For mid-term options (57 days), the same observations hold true. For long-term options, all the models yield a satisfactory fit (except BGM for NFLX stock), and the HKDE model displays no decisive advantage over its competitors for such tenors, though it still offers a very precise fit. However, as the short-expiry case is generally the most difficult for any model to handle satisfactorily, this highlights a significant advantage of HKDE.


{
\begin{figure}
    \centering
    \begin{subfigure}{0.30\textwidth}
      \centering
      \includegraphics[width=\linewidth]{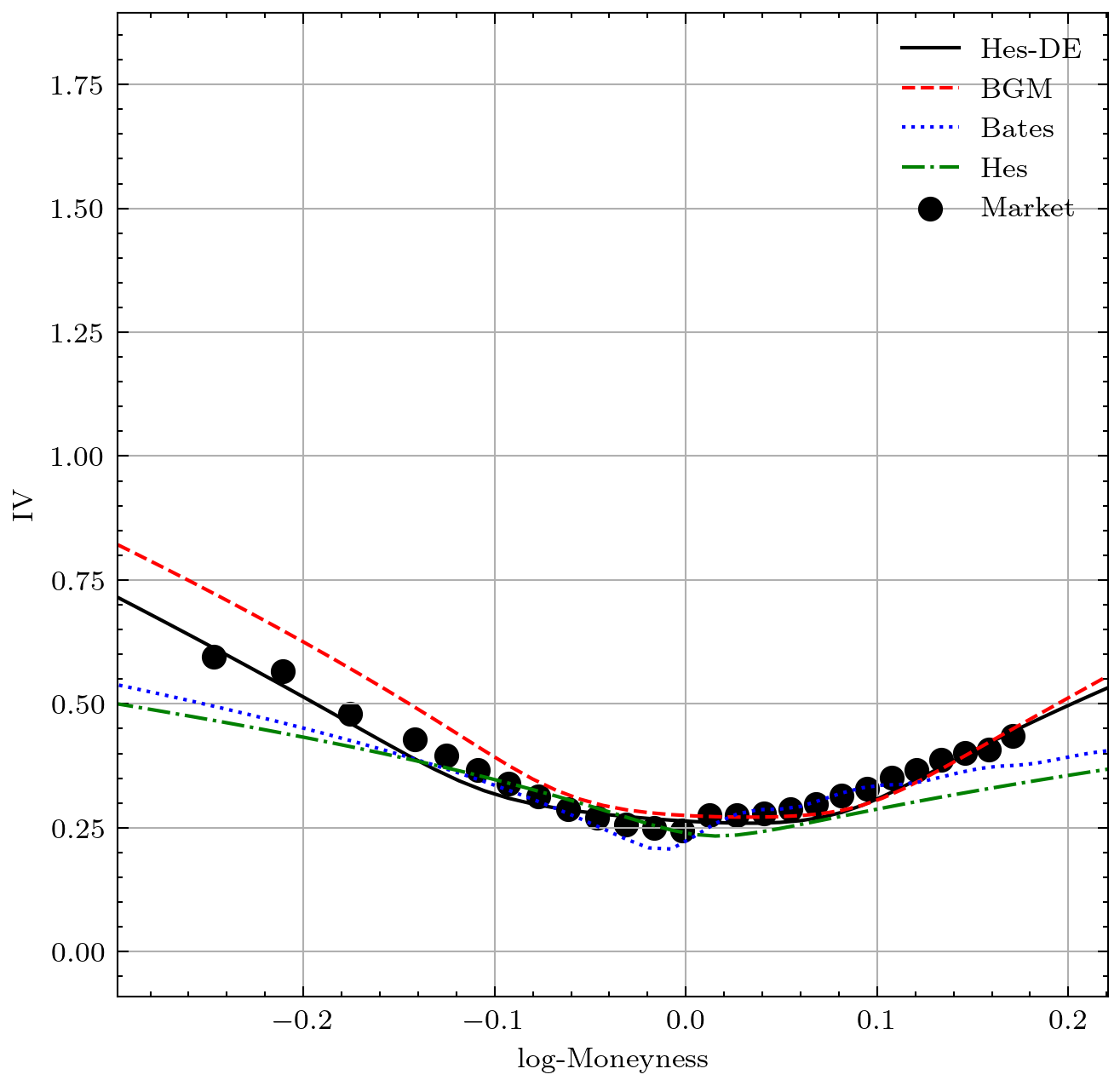}
      \caption{\footnotesize{AMZN - 8 days}}
    \end{subfigure}
    \begin{subfigure}{0.30\textwidth}
      \centering
      \includegraphics[width=\linewidth]{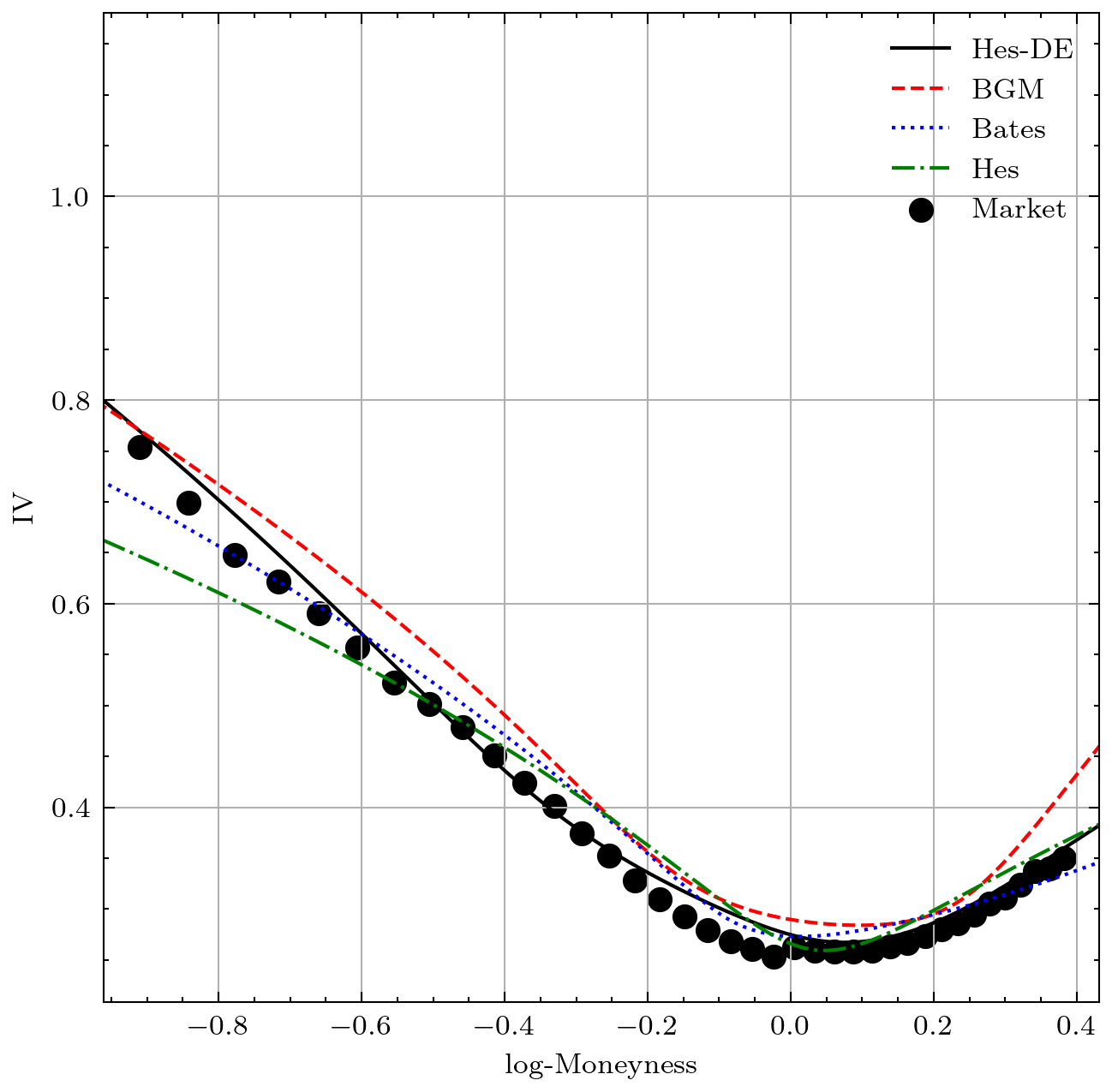}
      \caption{\footnotesize{AMZN - 57 days}}
    \end{subfigure}
    \begin{subfigure}{0.30\textwidth}
      \centering
      \includegraphics[width=\linewidth]{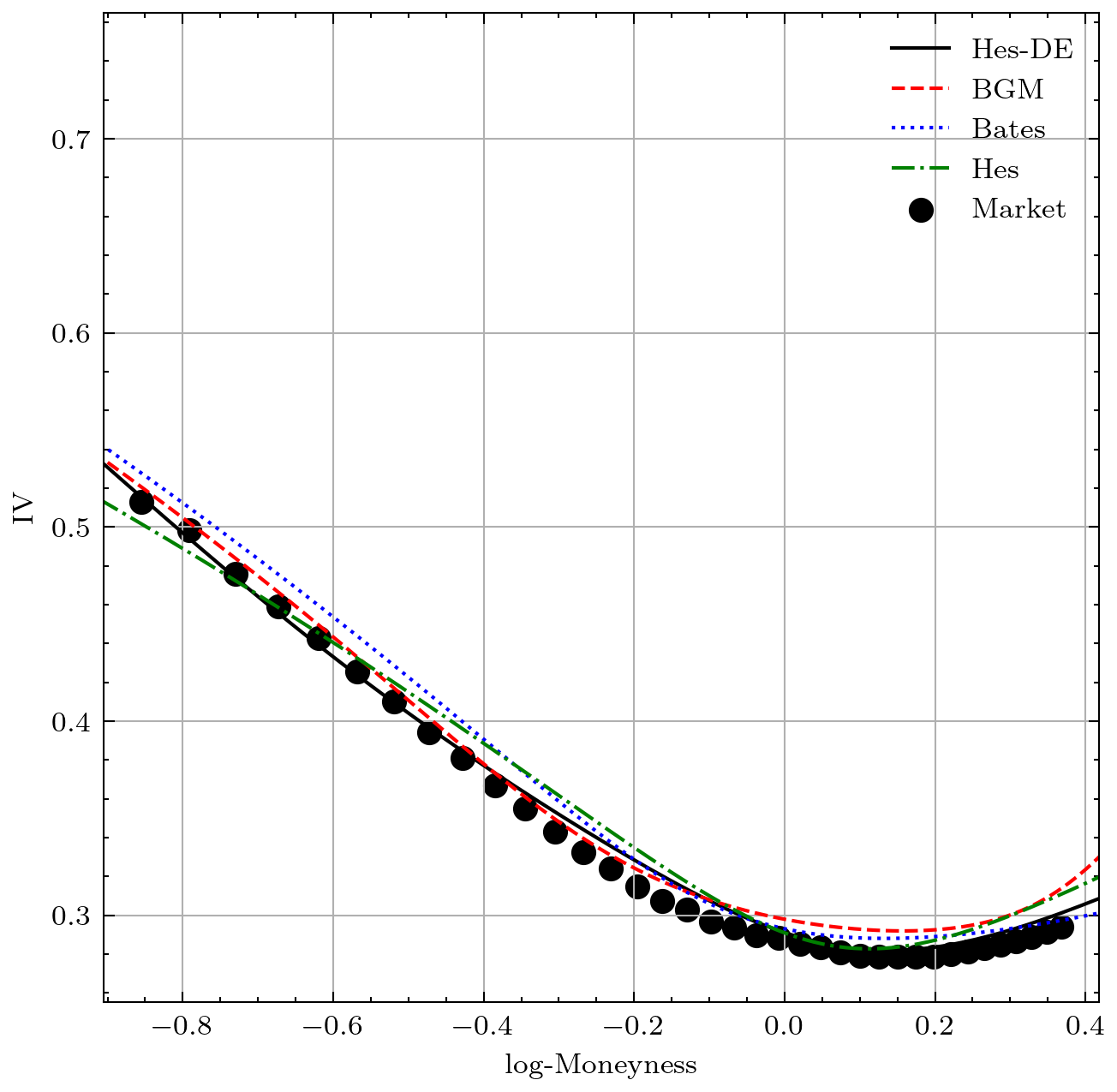}
      \caption{\footnotesize{AMZN - 148 days}}
    \end{subfigure}\\
    \begin{subfigure}{0.30\textwidth}
      \centering
      \includegraphics[width=\linewidth]{Figures/ImpliedVol/AMZN/AMZN_8.png}
      \caption{\footnotesize{NFLX - 8 days}}
    \end{subfigure}
    \begin{subfigure}{0.30\textwidth}
      \centering
      \includegraphics[width=\linewidth]{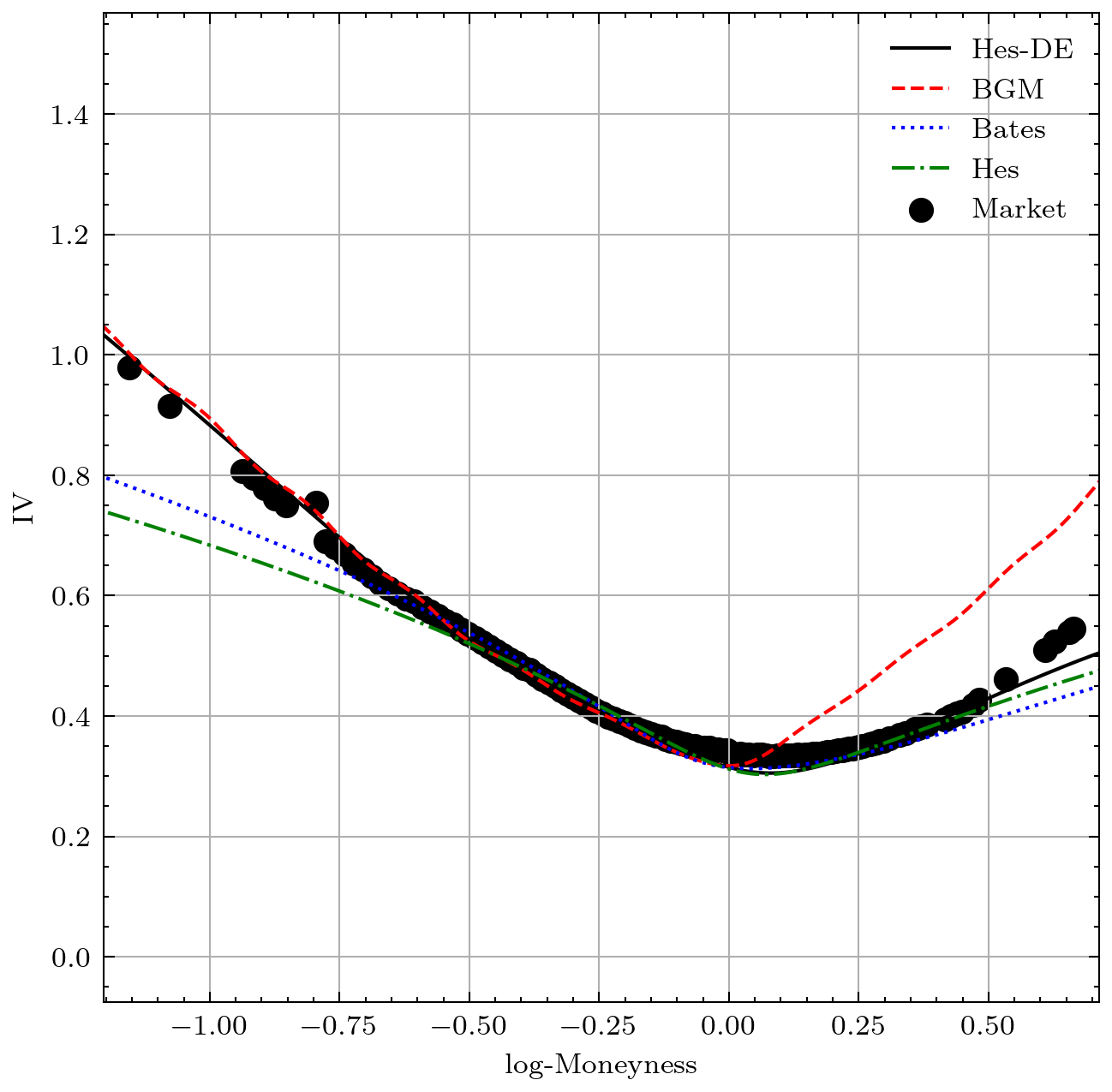}
      \caption{\footnotesize{NFLX - 57 days}}
    \end{subfigure}
    \begin{subfigure}{0.30\textwidth}
      \centering
      \includegraphics[width=\linewidth]{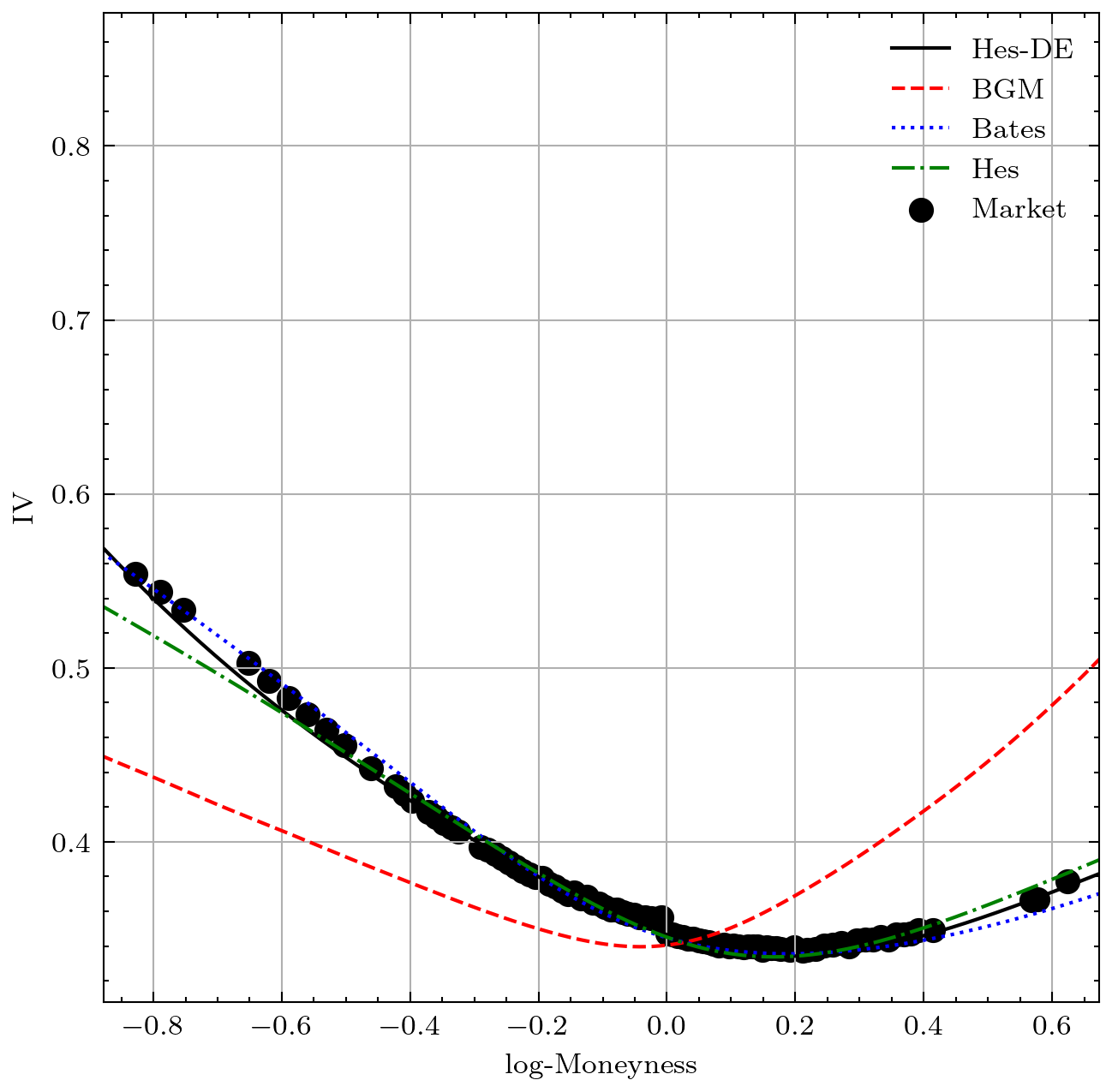}
      \caption{\footnotesize{NFLX - 148 days}}
    \end{subfigure}\\
    \begin{subfigure}{0.30\textwidth}
      \centering
      \includegraphics[width=\linewidth]{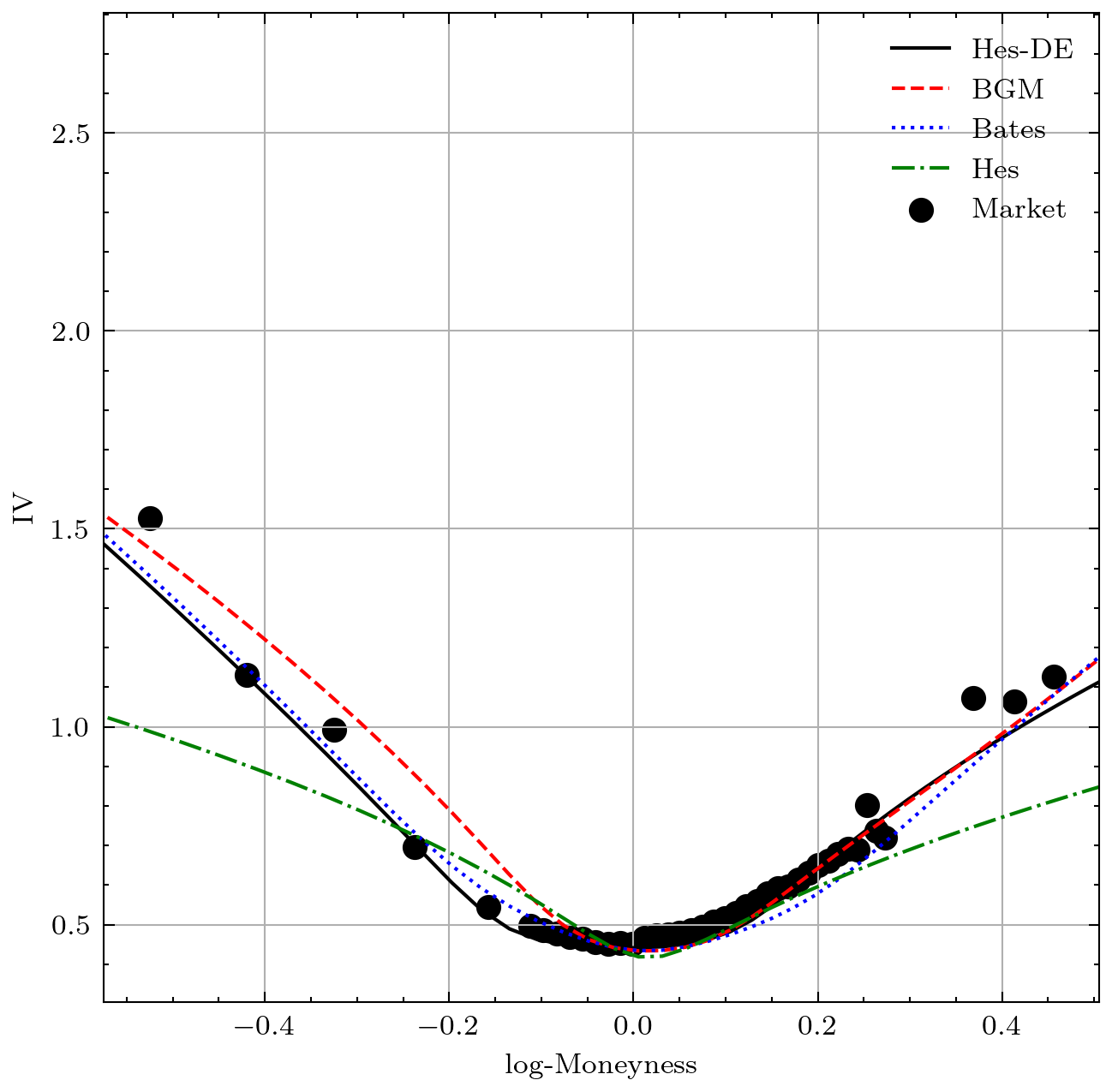}
      \caption{\footnotesize{SHOP - 8 days}}
    \end{subfigure}
    \begin{subfigure}{0.30\textwidth}
      \centering
      \includegraphics[width=\linewidth]{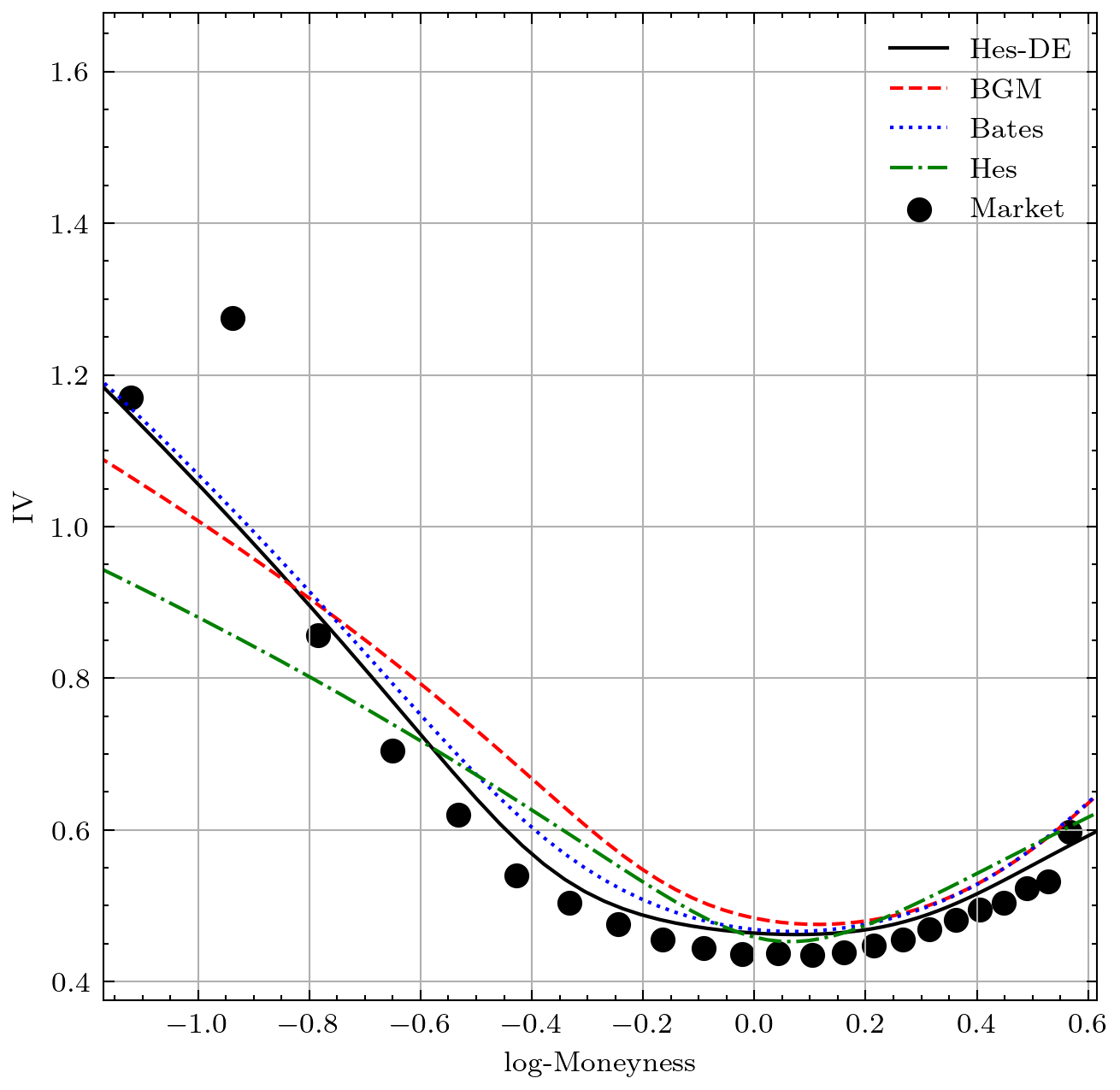}
      \caption{\footnotesize{SHOP - 57 days}}
    \end{subfigure}
    \begin{subfigure}{0.30\textwidth}
      \centering
      \includegraphics[width=\linewidth]{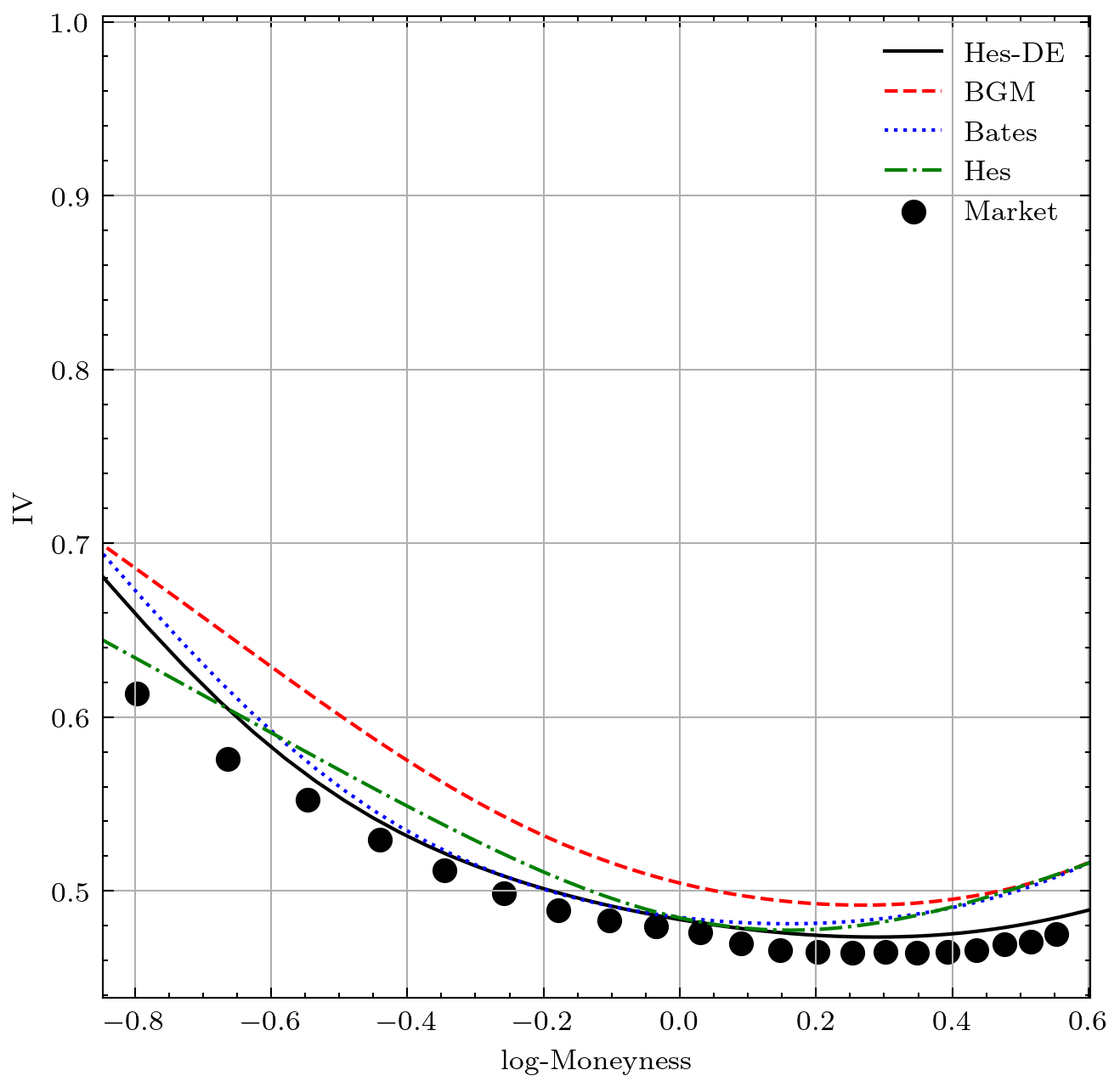}
      \caption{\footnotesize{SHOP - 148 days}}
    \end{subfigure}\\
    \begin{subfigure}{0.30\textwidth}
      \centering
      \includegraphics[width=\linewidth]{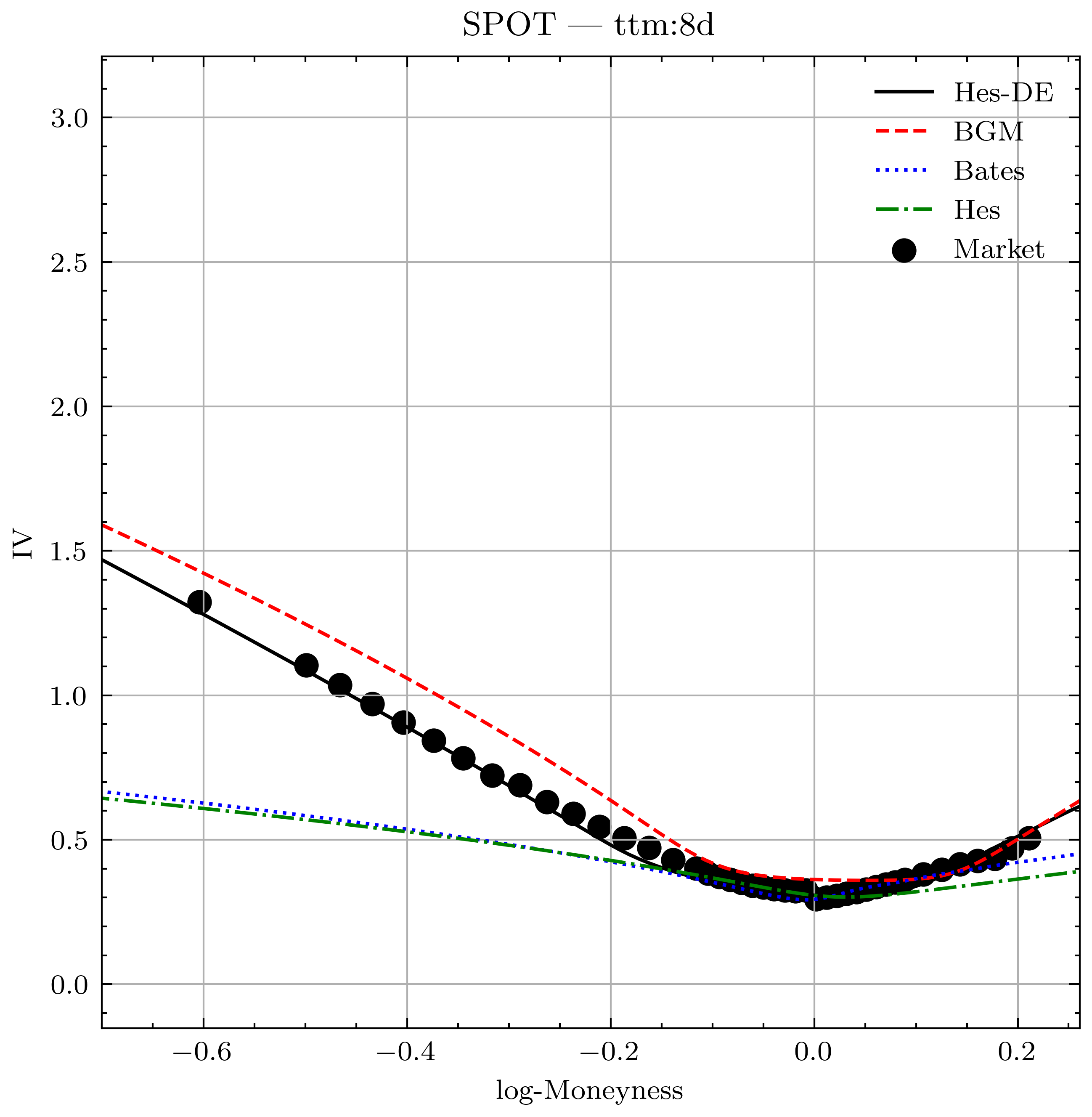}
      \caption{\footnotesize{SPOT - 8 days}}
    \end{subfigure}
    \begin{subfigure}{0.30\textwidth}
      \centering
      \includegraphics[width=\linewidth]{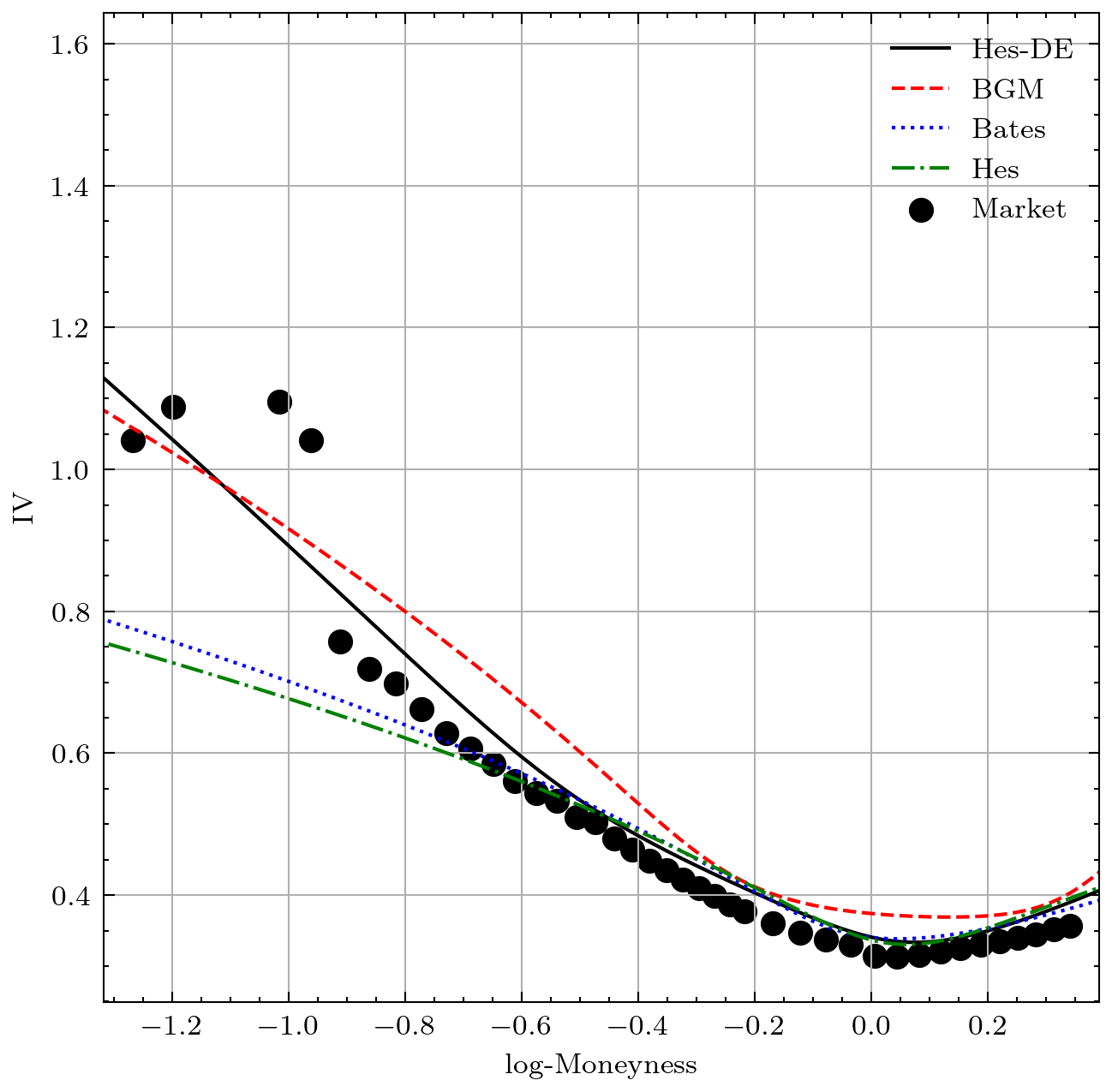}
      \caption{\footnotesize{SPOT - 57 days}}
    \end{subfigure}
    \begin{subfigure}{0.30\textwidth}
      \centering
      \includegraphics[width=\linewidth]{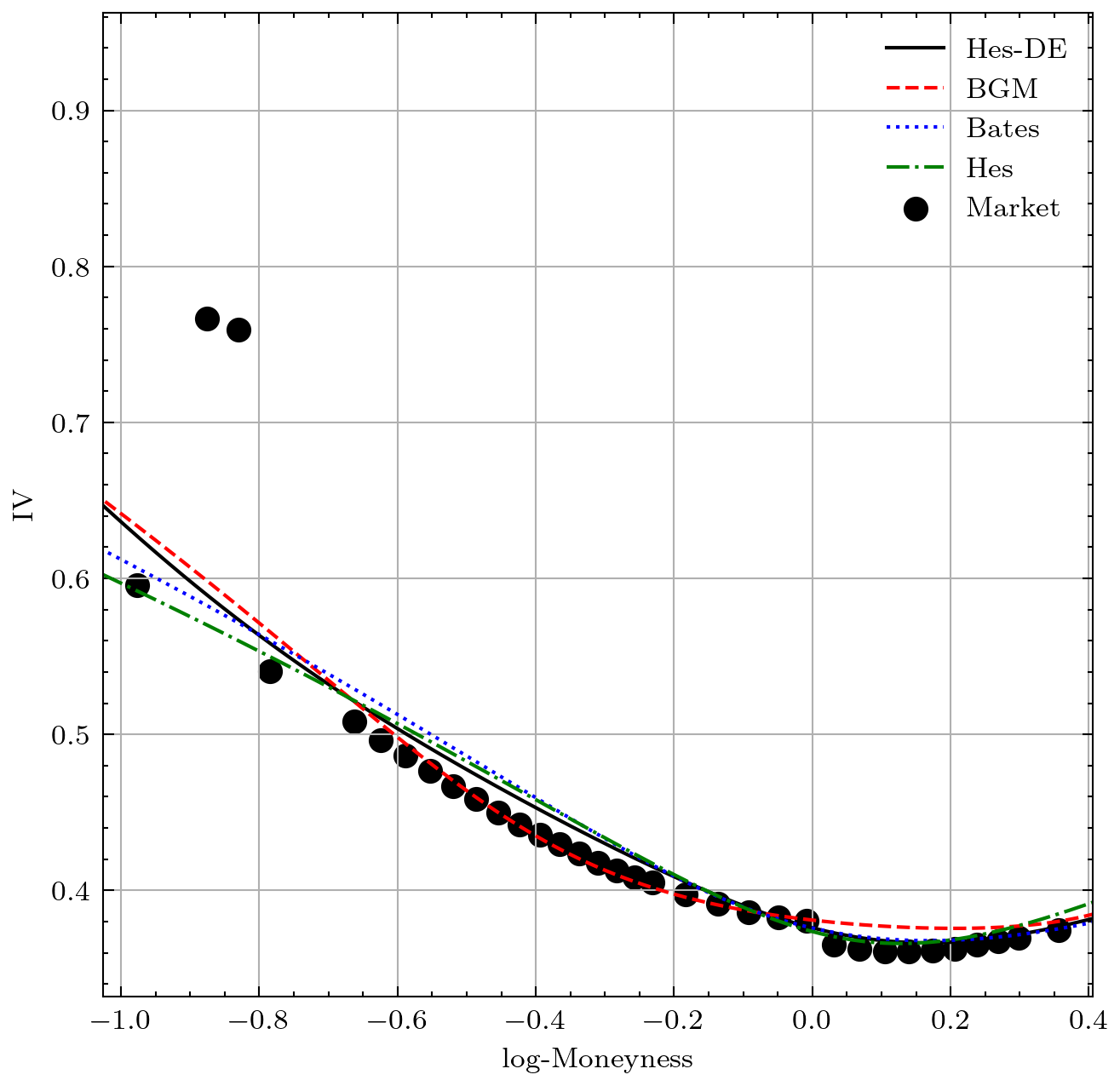}
      \caption{\footnotesize{SPOT - 148 days}}
    \end{subfigure}\\
    \caption{\footnotesize{
    Implied volatilities for AMZN, NFLX, SHOP and SPOT under the four tested models. The curves represent cross-sections of the implied volatility surfaces.}
    }
    \label{fig:ImpliedVolSmile}
\end{figure}
}


\section{Exotic Option Pricing}\label{sec:Exotics}

This section presents the pricing of four exotic options under the four calibrated models, and seeks to highlight the differences in model prices and the resulting sensitivity of some common exotics to the calibrated models. The contracts analyzed are arithmetic Asian options, cliquets, options on discretely sampled realized variance, and Barrier options. 
 The pricing algorithms are based on the PROJ valuation framework, which has been optimized and implemented into the open source {\sffamily fypy} library in {\sffamily Python}.  These algorithms are also available in {\sffamily Matlab}.\footnote{See \url{https://github.com/jkirkby3/PROJ\_Option\_Pricing\_Matlab}.}
 The particular algorithms and analyses for each type of contract are detailed in \cite{kirkby2016efficient, unifiedBarrierBermudanStoKirkby,cui2017equity, KirkbyDVOptionSto}. Although other methods can be applied, PROJ is recognized as one of the fastest and most accurate techniques available for vanilla and exotic options.\\
Pricing will be conducted using the calibrated surface parameters specific to each market. The calibrated values are the ones presented in Table\ref{table:CalibParams}.
To ensure consistency and facilitate cross-asset comparisons, the following parameters will be standardized across all markets: spot price, maturity, number of monitoring points, interest rate, and dividend yield. 

\subsection{Recursive Return Derivatives}
{\color{black}
        {\color{black}
        \begin{table}[h!t!b!]
        \centering
        \begin{tabular}{lll}
        \toprule
        \textbf{Contract Type} & $A_M$ & $G(A_M)$ \\
        \midrule
        Realized Variance Swaps & $\displaystyle  \frac{1}{T}\sum_{m=1}^{M}\left( R_m \right)^2$ & $A_M-K$ \\
        Realized Variance Options & $\displaystyle  \frac{1}{T}\sum_{m=1}^{M}\left(e^{R_m} - 1 \right)^2$ & $(A_M-K)^+$ \\
        Cliquet & $\displaystyle \sum_{m=1}^M \max\left(F,\min\left(C,\exp(R_m) - 1\right) \right)$ &  $K \cdot\min\left(C_g, \max\left(F_g, A_M \right) \right)$ \\
        Asian Options & {\scriptsize $\displaystyle\frac{S_0}{M+1}\left(w_0 + \e^{R_1}\left(\dots\left(w_{M-1} + w_{M}\e^{R_M} \right)\right)\right)$} & $(A_M-K)^+$ \\
        \bottomrule
        \end{tabular}
        \caption{\footnotesize{Definitions of $A_M$ and $G(A_M)$ for ones of the most popular Recursive Return Derivatives.}}
        \label{tab:recursive_price}
        \end{table}
        
The Fourier-based pricing methodology considered in this work encompasses a wide range of exotic contract types. The first type can be categorized in terms of a generic recursive structure which they satisfy. We refer to these contracts as \emph{recursive return derivatives}, and they include many common exotics; see also \cite{leitao2021ctmc} who first provided this characterization.  Consider a fixed time horizon $[0,T]$ over which there are $M + 1$ monitoring dates, $0 = t_0 < t_1 < \cdots < t_M = T$.  We define the log-return $R_m$ between monitoring dates by 
		\begin{equation*}
			R_m:= \log\left(\frac{S_m}{S_{m-1}}\right), \quad S_m :=S(t_m), \quad m=1,...,M,
		\end{equation*}
		where $t_{m+1} - t_m$ is not required to be uniform, but we will assume for simplicity that $t_{m+1} - t_m = T / (M+1)$ is uniform. The recursive return derivatives satisfy a general sequence of equations
		\begin{equation}\label{eq:genRecur}
		\begin{aligned}
			&Y_1 : = w_M \cdot h(R_M) + \varrho_M \\
			&Y_m: =  w_{M-(m-1)} \cdot h(R_{M-(m-1)}) + g(Y_{m-1}) + \varrho_{M-(m-1)},\quad m=2,\ldots,M,
		\end{aligned}
		\end{equation}
		where $h,g:\mathbb R \rightarrow \mathbb R$ are continuous functions,  $\{w_m\}_{m=1}^M$ is a set of weights, and $\{\varrho_m\}_{m=1}^M$ is a set of shift parameters. This formulation includes all contracts of the form 
		\[
			 G\left(\sum_{m=1}^M w_m \cdot h(R_m) ; \Pi \right),
		\]
		where $\Pi$ is a set of generic contract parameters. For example, when $h(R_m) = \exp(R_m) - 1$ or $h(R_m) = R_m$, we recover two common varieties of realized variance derivatives in addition to cliquets/ratchets/equity indexed annuities. This formulation also covers more general iterated compositions of the form
		\[
		 G\left(\bigotimes_{m=1}^M h_m(R_m) ; \Pi \right) ,
		\]
		where $\bigotimes_{m=1}^M$ denotes the composition $h_1(R_1) \circ h_2(R_2) \circ \cdots \circ h_M(R_M)$, where summation is a special case. Important examples of recursive return derivatives are mentioned in Table \ref{tab:recursive_price}. We denote $A_M$ as the quantity of interest related to the underlying asset, and $G(A_M)$ as the claim's payoff. 

        The only necessary difference in the pricing procedure for contracts of this form is the specification of the terminal payoff.}

\subsubsection{Asian Option Prices}
In order to compute the prices of Arithmetic Asian options, we apply the APROJ Fourier algorithm of \cite{kirkby2016efficient} to price under the BGM model, and the one of \cite{kirkby2020efficient} to price under the Heston, Bates and HKDE models. Results are presented in Table \ref{table:AsianPrices}.

The Asian payoff (due to its averaging) is clearly smoother than the other claims mentioned in this article, resulting in option prices that are less sensitive to the underlying model. In particular, capturing the overall volatility level (especially near the money) is more important than the surface wings (i.e.~the tails of the transition density), resulting in less variation between different models, even across different model families. Consequently, the variance of the option prices is tightly contained across all underlyings. The highest percentage differences are consistently observed for OTM options. The most significant discrepancy is noted for the stock SHOP between the Heston and HKDE models, amounting to $\$0.41$, which represents a $13\%$ difference.  However, as a consequence we can expect tighter bid-ask spreads on these contracts, and even a $13\%$ difference in price could be significant, depending on the market.

\begin{table}[h!t!b!]
\centering 
\scalebox{.8}{
\begin{tabular}{ c| M{2.5cm} M{2.5cm} M{2.5cm} | M{2.5cm} M{2.5cm} M{2.5cm} }
\hline
\rule{0pt}{1.8\normalbaselineskip}\multirow{-2}{*}& &\vspace{\fill}\textbf{AMZN} \vspace{1mm}&& &\vspace{\fill}\textbf{SHOP} \vspace{1mm}& \\ \hline
\rule{0pt}{1.4\normalbaselineskip}\textbf{Model} &  \textbf{ITM} ($\$$)  & \textbf{ATM} ($\$$) & \textbf{OTM} ($\$$) &  \textbf{ITM} ($\$$) & \textbf{ATM} ($\$$) & \textbf{OTM} ($\$$) \\ \hline%

  \rule{0pt}{2.5ex}   BGM     &   31.17    &    7.90     &    0.67    &  32.19    &  12.22    &    3.48  \\
    Heston     &   31.18    &    7.92     &    0.79    &  32.06     &   12.13    &    3.66 \\
    HKDE     &   31.21    &    8.00     &    0.76    &  32.01     &   11.95    &     3.25 \\
  Bates     &   31.17    &    7.91     &    0.73    &  32.05     &   12.03    &     3.60  \\
  \hline

\rule{0pt}{1.8\normalbaselineskip}\multirow{-2}{*}& &\vspace{\fill}\textbf{NFLX} \vspace{1mm}&& &\vspace{\fill}\textbf{SPOT} \vspace{1mm}& \\ \hline
\rule{0pt}{1.4\normalbaselineskip}\textbf{Model} &  \textbf{ITM} ($\$$) & \textbf{ATM} ($\$$) & \textbf{OTM} ($\$$)&  \textbf{ITM} ($\$$) & \textbf{ATM} ($\$$) & \textbf{OTM} ($\$$) \\ \hline%

  \rule{0pt}{2.5ex}   BGM     &   31.31    &    8.94     &    1.22    &  31.42     &   9.70    &     1.71  \\
    Heston     &   31.33   &    9.06     &    1.32    &  31.50     &   9.73    &     1.83  \\
    HKDE     &   31.34    &    9.09     &    1.41    &  31.44     &   9.73    &     1.72  \\
  Bates     &  31.29    &    9.01     &    1.22    &  31.45    &   9.68    &     1.71  \\

  \hline

\end{tabular}
}
\caption{\color{black}\footnotesize{Prices for Arithmetic Asian Call options, computed using calibrated parameters and APROJ. \\ Spot = \$100, ITM strike= \$70, OTM strike= \$130, maturity= 9 months ($T=9/12)$, monitoring points = 40, risk-free rate= $5\%$, dividend yield= $0\%$.}} 
\label{table:AsianPrices}
\end{table}


\subsubsection{Discrete Variance Options Prices}

Discrete Variance options have been priced using the methodology from \cite{KirkbyDVOptionSto}, and the results are presented in Table \ref{table:dvOptionPrices}. We provide the prices for Discrete Variance options at three strike levels: $K\in \{0.01,0.03,0.05\}$. The prices are given to the third decimal place to highlight the differences between the models. 

For all tickers, we observe the highest differences, in percentage, for the strike $K=0.05$. This derivative exhibits significant sensitivity to the underlying model. The underlying SPOT exhibits the most significant discrepancies, showing a $\$0.094$ difference between Bates and BGM, which is a significant $81\%$ difference. We can observe that of the four model prices for SPOT, the Bates model is the clear outlier. Following SPOT, SHOP shows notable variations, with a $37\%$ difference between HKDE and Heston. This example clearly highlights the importance of proper model calibration, as well as the need to use multiple models to assess the reasonableness of calibrated prices. We also note that a L\'evy model such as BGM should not be seriously considered for pricing realized-variance derivatives, as it has no mechanism to model the ``vol-of-vol", and is thus ill-suited for capturing and risk-managing variance sensitive products. By contrast, stochastic volatility models such as Heston and its jump extensions were specifically designed for this purpose.

\begin{table}[h!t!b!]
\centering 
\scalebox{.8}{
\begin{tabular}{ c| M{2.5cm} M{2.5cm} M{2.5cm} | M{2.5cm} M{2.5cm} M{2.5cm} }
\hline
\rule{0pt}{1.8\normalbaselineskip}\multirow{-2}{*}& &\vspace{\fill}\textbf{AMZN} \vspace{1mm}&& &\vspace{\fill}\textbf{SHOP} \vspace{1mm}& \\ \hline
\rule{0pt}{1.4\normalbaselineskip}\textbf{Model} &  $K=0.01 $  & $K=0.03$ & $K=0.05$ &  $K=0.01$  & $K=0.03$ & $K=0.05$ \\ \hline%

\rule{0pt}{2.5ex}   BGM       &   0.089    &    0.070     &0.052            &  0.266    & 0.247 &0.228      \\
    Heston                    &   0.091    &  0.072 &0.056            &  0.246     & 0.227 & 0.208     \\
    HKDE                      &   0.099    &  0.079 &0.062            &  0.322     & 0.303&  0.284   \\
   Bates                       &   0.097   &  0.078 &0.062            &  0.298     &0.280 &  0.261    \\
  \hline

\rule{0pt}{1.8\normalbaselineskip}\multirow{-2}{*}& &\vspace{\fill}\textbf{NFLX} \vspace{1mm}&& &\vspace{\fill}\textbf{SPOT} \vspace{1mm}& \\ \hline
\rule{0pt}{1.4\normalbaselineskip}\textbf{Model} &  $K=0.01$  & $K=0.03$ & $K=0.05$ &  $K=0.01$  & $K=0.03$ & $K=0.05$ \\ \hline%

  \rule{0pt}{2.5ex}   BGM     &   0.127   & 0.108 & 0.089        &  0.154     & 0.135 &  0.116      \\
    Heston                      &   0.128     & 0.109 &  0.091         &  0.165   & 0.146 &  0.128       \\
    HKDE                    &   0.128     & 0.109 &  0.090         &  0.157    &0.138 &   0.120       \\
  Bates                     &  0.140      &  0.121&0.103         &  0.232    &  0.218 &0.210        \\

  \hline

\end{tabular}
}
\caption{\color{black}\footnotesize{Prices $(\$)$ for Variance Call options, computed using calibrated parameters. \\ Maturity= 1 year ($T=1)$, monitoring points = 40, risk-free rate= $5\%$, dividend yield= $0\%$.  }} 
\label{table:dvOptionPrices}
\end{table}

\subsubsection{Cliquet Options Prices}
The prices for Cliquet options are detailed in Table \ref{table:cliquetOptionPrices}, priced using the methodology in \cite{cui2017equity}. We present the prices for Cliquet options at three strike levels: $K\in \{0.5, 1, 1.5\}$. Similarly to Discrete Variance options, prices are shown to the third decimal place to emphasize the distinctions between the models. 
As anticipated, price differences for Cliquet options are more moderate compared to Variance Options, as the cap and floors effectively limit the extent of tail differences between the models. This is of course dependent on how wide/symmetric the caps/floor are set. The most significant price variations are observed in SHOP and AMZN. For both tickers, the highest price differences occur at the strike $K=1.5$, with the Heston model providing the lowest prices. Specifically, in SHOP, the HKDE model shows the highest price difference, with a 17\% spread between the models, while in AMZN, the BGM model shows a 15\% spread.  The observation that Heston produces lower prices than the models with a jump component highlights the importance of modeling jumps for return-sensitive assets such as cliquets, which can be very sensitive to the occurrence of price jumps. As the contracts ``reset" with each monitoring point, it allows for multiple jumps during the life of the contract to produce outsized returns, as the cap permits.
\begin{table}[h!t!b!]
\centering 
\scalebox{.8}{
\begin{tabular}{ c| M{2.5cm} M{2.5cm} M{2.5cm} | M{2.5cm} M{2.5cm} M{2.5cm} }
\hline
\rule{0pt}{1.8\normalbaselineskip}\multirow{-2}{*}& &\vspace{\fill}\textbf{AMZN} \vspace{1mm}&& &\vspace{\fill}\textbf{SHOP} \vspace{1mm}& \\ \hline
\rule{0pt}{1.4\normalbaselineskip}\textbf{Model} &  $K=0.5$  & $K=1$ & $K=1.5$ &  $K=0.5$  & $K=1$ & $K=1.5$ \\ \hline%

\rule{0pt}{2.5ex}   BGM       &   0.403    &  0.806 &  1.210            &  0.458     & 0.916 &  1.374      \\
    Heston                    &   0.351    &  0.702 &  1.053            &  0.419     & 0.838 &  1.257     \\
    HKDE                      &   0.392    &  0.784 &  1.176            &  0.488     & 0.977 &  1.466   \\
   Bates                      &    0.398   &  0.796 &  1.194            &  0.462     &0.925  &  1.387    \\
  \hline

\rule{0pt}{1.8\normalbaselineskip}\multirow{-2}{*}& &\vspace{\fill}\textbf{NFLX} \vspace{1mm}&& &\vspace{\fill}\textbf{SPOT} \vspace{1mm}& \\ \hline
\rule{0pt}{1.4\normalbaselineskip}\textbf{Model} &  $K=0.5$  & $K=1$ & $K=1.5$ &  $K=0.5$  & $K=1$ & $K=1.5$ \\ \hline%

  \rule{0pt}{2.5ex}   BGM     &   0.440   & 0.880  &  1.320          &  0.457     & 0.914  &  1.372      \\
    Heston                    &   0.393   & 0.786  &  1.180         &  0.410     & 0.820  &  1.230       \\
    HKDE                      &  0.400    & 0.801  &  1.202        &  0.428     &0.856   &   1.284       \\
  Bates                       &  0.410    &  0.820 &1.230         &  0.431     &  0.863 &1.295        \\

  \hline

\end{tabular}
}
\caption{\color{black}\footnotesize{Prices $(\$)$ for Cliquet options, computed using calibrated parameters. \\ Maturity= 1 year ($T=1)$, monitoring points ($M$) = 40, risk-free rate= $5\%$, dividend yield= $0\%$, $C=0.06$, $F=0.01$, $C_g = 0.75\times M\times C$, $F_g = 1.25\times M\times F$.  }} 
\label{table:cliquetOptionPrices}
\end{table}

\subsection{Barrier Options}
{\color{black}We next consider the case of a barrier option, which is a special case of the occupation time derivatives.
A \emph{knock-out} barrier option awards the contract purchaser with payout $G(S_T)=G(S_{t_M})$, as long as $S(t_m)$ is never observed outside of a specified continuation region $\mathcal C$ for any $t_m\in \mathcal T_m$, where $\mathcal T_m$ denotes the set of monitoring dates. If we define 
\begin{align*}
S^{max}_M &:= \max\{S_m : 1\leq m\leq M\}\\
S^{min}_M &:= \min\{S_m : 1\leq m\leq M\}, 
\end{align*}
the three main varieties of \emph{knock-out} options can be described as: an \emph{up-and-out} contract with \\ 
$\mathcal C =[0,U]$ paying at maturity $G(S_M)\bm{1}_{\{S^{max}_M\leq U\}}$, a \emph{down-and-out} contract with $\mathcal C = [L,\infty)$ paying $G(S_M)\bm{1}_{ \{ L\leq S^{min}_M\}}$, and a \emph{double-barrier} contract with $\mathcal C = [L,U]$  paying $G(S_M)\bm{1}_{\{L\leq S^{min}_M\}}\bm{1}_{\{S^{max}_M\leq U\}}$.  The value of such contracts is defined with respect to the risk-neutral measure as
\begin{equation*}
\mathcal V(S_0) = e^{-rT}\mathbb E\Big[g(S_M)\prod_{1\leq m\leq M}\bm{1}_{\{S_m\in \mathcal C\}} |S_0 \Big] = e^{-rT}\mathbb E\left[G(S_M)\bm{1}_{\{\tau^{\mathcal C} \not\leq T\}}|S_0 \right] ,
\end{equation*}
where $\tau^{\mathcal C} := \inf_{m\geq 1}\{t_m: S(t_m) \notin \mathcal C\}$. \emph{Knock-in} options are contracts whose payoff is zero unless the underlying $S(t_m)$ is observed outside of $\mathcal C$ for some $t_m\in \mathcal T_m$. Their terminal payoff is of the form $G(S_M)\bm{1}_{\{\tau^{\mathcal C} \leq T\}}$, and they are priced using parity relationships.}  

{\color{black}
\subsubsection{Barrier Option Prices}

To price barrier options, we utilize the technique developed in \cite{unifiedBarrierBermudanStoKirkby}, which enables the pricing of both Barrier and Bermudan options under stochastic volatility and jump models. For L\'evy models (in this case BGM) we use the methodology of \cite{kirkby2017robust}. Generally speaking, the strong path dependency of barrier options leads market practitioners to price them under stochastic (and sometimes stochastic local) volatility models.}

\begin{table}[h!t!b!]
\centering 
\scalebox{.8}{
\begin{tabular}{ c| M{2.5cm} M{2.5cm} M{2.5cm} | M{2.5cm} M{2.5cm} M{2.5cm} }
\hline
\rule{0pt}{1.8\normalbaselineskip}\multirow{-2}{*}& &\vspace{\fill}\textbf{AMZN} \vspace{1mm}&& &\vspace{\fill}\textbf{NFLX} \vspace{1mm}& \\ \hline
\rule{0pt}{1.4\normalbaselineskip}\textbf{Model} &  \textbf{ITM} ($\$$)  & \textbf{ATM} ($\$$) & \textbf{OTM} ($\$$) &  \textbf{ITM} ($\$$) & \textbf{ATM} ($\$$) & \textbf{OTM} ($\$$) \\ \hline%

  \rule{0pt}{2.5ex}  BGM    &    18.76    &    4.46   &    0.12    &      15.17     &    3.42    &     0.09     \\
   Heston    &    19.54    &    4.98    &     0.16    &     15.38     &    3.77    &     0.13   \\
   HKDE    &    18.75    &   4.62    &     0.13   &      15.15     &    3.58    &     0.11   \\
 Bates    &    19.32    &    4.90    &    0.15   &      15.13     &    3.64    &     0.12   \\

  \hline

\rule{0pt}{1.8\normalbaselineskip}\multirow{-2}{*}& &\vspace{\fill}\textbf{SHOP} \vspace{1mm}&& &\vspace{\fill}\textbf{SPOT} \vspace{1mm}& \\ \hline
\rule{0pt}{1.4\normalbaselineskip}\textbf{Model} &  \textbf{ITM} ($\$$)  & \textbf{ATM} ($\$$) & \textbf{OTM} ($\$$) &  \textbf{ITM} ($\$$) & \textbf{ATM} ($\$$) & \textbf{OTM} ($\$$) \\ \hline%

  \rule{0pt}{2.5ex}     BGM    &    8.91    &   1.99    &         0.06    &    13.08    &     2.84  & 0.08    \\
   Heston    &    9.14   &    2.05   &     0.07   &      13.73     &    3.32    &     0.11  \\
   HKDE    &    8.30    &    1.78    &     0.06    &      13.17     &    3.11   &     0.23   \\
 Bates    &    8.93    &    1.88    &     0.06    &      13.38     &    3.20    &     0.10   \\

  \hline

\end{tabular}
}
\caption{\color{black}\footnotesize{Prices for barrier up-and-out call options, computed using calibrated parameters. \\ Spot = \$100, ITM strike = \$70, OTM strike = \$130, upper barrier = $\$140$, maturity = 1 year ($T=1)$, monitoring points = 40, risk-free rate = $5\%$, dividend yield = $0\%$.}} 
\label{table:BarrierPrices}
\end{table}

{\color{black}
Table \ref{table:BarrierPrices} presents the prices of barrier up-and-out call options for each of the underlyings considered in this article. At first glance, the price differences between ITM, ATM, and OTM options for each underlying and each model appear consistent.
However, this specific derivative is the one that exhibits the highest percentage differences observed. For instance, for the ticker SPOT, the OTM price quoted by HKDE is almost three times that given by BGM.
This is not surprising: since barrier options knock-out if there are significant fluctuations in the underlying asset, they are particularly affected by the volatility wings implied by the model.  We also note that the Heston model consistently produces the highest prices; as Heston is the only model without jumps, this reflects the reduced risk that model ascribes to triggering the knock-out event. It also suggests that a jump component could be important for ensuring accurate pricing of knock-out contracts in markets where jumps are routinely observed.
As suggested in \cite{carr2010class, guo2021optimal}, doing a calibration including both vanilla and exotic options (such as barriers) could provide more accurate prices. 
}

\section{Conclusion}\label{sec:Conclu}
In this work we study a natural generalization of the Heston model via the introduction of a jump-diffusion component in the asset price dynamics. This model is called the Heston-Kou double exponential (HKDE) model. By featuring an asymmetric distribution of the jump sizes, the HKDE model addresses a well-known limitation of other generalizations (such as the Bates model) that are based on symmetric jump distributions, and allows for a more flexible behavior of the volatility smile. The improvement is particularly noticeable for short term options, where in our study the HKDE model consistently produces an excellent fit to the market smile for each underlying, while the Heston and Bates smiles tend to flatten out for deep ITM and OTM strikes, failing to adequately capture the wings of the implied volatility surface.

We have also provided a detailed impact analysis of the model's parameters over the shape of the implied volatility smile, as well as a first study of the model's behavior in pricing various options. This has been achieved by leveraging the PROJ method to evaluate several exotic payoffs (arithmetic Asian, barrier, Discrete Variance, cliquet) under HKDE dynamics.
This clearly demonstrates the tractability of the model via classical Fourier pricing techniques, as well as with their most modern refinements.
The experimental findings in this work demonstrate the potential of the HKDE model for practical applications. Natural extensions could include considering other distributions for the jumps log size and developing improved sampling algorithms for the model, as well as conducting empirical studies on the HKDE model with double-factor dynamics for the volatility process, or with the inclusion of a fractional Brownian motion.  The pricing of alternative exotic derivatives and calibration to other markets is also of interest.


\vspace{0.2cm}

\footnotesize
\bibliographystyle{plain} 
\bibliography{biblio}

\normalsize

\appendix


\section{Implementation Details}\label{app:implementation}
\subsection{Projection Coefficients}\label{sect:ProjCoeffs}
Here we describe the algorithm for determining the orthogonal projection coefficients of the density $f(x)$ onto the cubic B-spline basis, as discussed in Section \ref{sect:biorproj}. For more details, we refer the reader to \cite{kirkby2015efficient, kirkby2017robust}. The inputs are: $N$, $a$, and $\phi(\xi)$.
Recall the closed-form representation
\begin{equation}\label{eq: betadef2}
\beta_{a,n}:=\frac{a^{-1/2}}{\pi}\Re\left[\int_0^\infty \exp({-\mi x_n\xi}) \cdot \phi(\xi) \widehat{\widetilde \varphi}\Big(\frac{\xi}{a}\Big)d\xi\right].
\end{equation}
 We approximate the integral in \eqref{eq: betadef2} using the trapezoid rule, which is shown in \cite{kirkby2015efficient} to converge exponentially to the true coefficient values.
We define an $N$-point frequency grid
\begin{equation}\label{eq: frequencygridexotic}
\Delta_\xi = 2 \pi a/N, \qquad \xi_n = (n-1)\Delta_\xi, \quad n=1,...,N,
\end{equation}
where $\Delta_\xi$ is chosen according to the Nyquist frequency.
After applying the trapezoidal rule to \eqref{eq: betadef2} at grid points $\xi_n$, and collecting constants in the term $\Upsilon_{a,N}$, the coefficients $a^{1/2}\Upsilon_{a,N}\cdot\bar \beta_{a,k} \approx\beta_{a,k} $
 are found using
\begin{equation}\label{eq: betaj}
\left\{\bar \beta_{a,k}\right\}_{k=1}^N := \Re\left\{\mathcal D\left\{\textbf{H}\right\}\right\}, \quad \mathcal D_k\left\{\textbf{H}\right\} = \sum_{n=1}^N e^{-\mi \frac{2\pi}{N}(n-1)(k-1)} H_n, \quad k = 1,...,N,
\end{equation}
where $\mathcal D$ is the discrete Fourier transform (DFT).  The DFT input vector $\textbf{H}=\{H_n\}_{n=1}^N$ is defined as
\begin{equation}\label{eq:OriginalHDef}
H_1:=\phi(0)/32 a^4, \qquad H_n:=\phi(\xi_n)\cdot \zeta_n \cdot \exp(-\mi\xi_n\cdot x_1), \quad n\geq 2,
\end{equation}
where
\begin{equation}\label{eq: zetajeqexotic}
\zeta_n := \frac{2520(\sin(\xi_n/(2a))/\xi_n)^4 }{1208 + 1191 \cos(\xi_n/a) + 120 \cos(2\xi_n/a) + \cos (3\xi_n/a)}, \quad n\geq 2.
\end{equation}
The term $\zeta_n$ is simply a scaled version of the \emph{cubic} dual basis generator $ \widehat{\widetilde \varphi}(\xi/a)$ evaluated at $\xi_n$.
The coefficients can then be recovered at a cost of $\mathcal O(N \log_2(N))$ using the fast Fourier transform (FFT).  More generally, the dual generator ${\widetilde\varphi}^{[p]}(x)$ for a $pth$ order B-spline basis admits a closed-form Fourier transform which can be derived using
\[
\widehat{\widetilde\varphi}^{[p]}(\xi)=\left(\frac{\sin(\xi/2)}{(\xi/2)}\right)^{p+1}\left( \int_{-\frac{p+1}{2}}^{\frac{p+1}{2}} \varphi^{[p]}(x)^2dx  + 2\sum_{k=1}^{p+1}\cos(k\xi) \int_{-\frac{p+1}{2}}^{\frac{p+1}{2}} \varphi^{[p]}(x)\varphi^{[p]}(x-k)dx \right)^{-1}.
\]
 See \cite{kirkby2015efficient}  for more details.

\section{PROJ exotic pricing \textit{vs.} Monte-Carlo exotic pricing}

{
In this appendix, we wish to briefly justify our choice of choosing the PROJ method for calibration and exotic option pricing, by illustrating its efficiency on some examples; let us note that this efficiency has already been well studied and documented in the literature (see references below). We thus provide a few numerical comparisons between the PROJ method and Monte-Carlo procedure, for two different exotic derivatives, under HKDE dynamics. For Asian options, it can be observed that PROJ execution time is about $10^3$ times faster than Monte Carlo pricing whatever the moneyness, and 10 times faster for barrier options, confirming the results already observed in \cite{kirkby2020efficient,unifiedBarrierBermudanStoKirkby}. Detailed numerical results are displayed in tables \ref{tab:comp-barrier} and \ref{tab:comp-asian}. For a completed study of the performance of the PROJ pricing method under models featuring stochastic volatility and jumps and for the whole set of exotic instruments under consideration in the present paper, we invite the reader to refer to:
\begin{itemize}
    \item \cite{kirkby2020efficient} for Asian option;
    \item \cite{KirkbyDVOptionSto} for variance options;
    \item \cite{unifiedBarrierBermudanStoKirkby} for barrier options (Bermudan are also studied in this paper);
    \item \cite{cui2017equity} for cliquet-style options.
\end{itemize}
All these articles clearly demonstrate the ability of the PROJ method to handle exotic option pricing efficiently, and to bring significant improvements in terms of computation time and precision when compared to  Monte-Carlo pricing and even to other pricing methods when available. 
}

\begin{table}[h!]
\centering
\begin{tabular}{|p{1cm}|p{1.5cm}|p{1.5cm}|p{1.5cm}|p{2cm}|p{2.5cm}|p{1.9cm}|p{1.8cm}|}
\hline
\textbf{Strike} &
\textbf{Points} & \textbf{Ticker} & \textbf{PROJ} & \textbf{MC} & \textbf{Confidence MC (95\%)} & \textbf{Time PROJ (s)} & \textbf{Time MC (s)} \\ \hline\hline

 70 & 2                         & AMZN           & 32.30265           & 32.34305         & $\pm$ 0.14224                & 0.0376                 & 23.4025              \\ \cline{3-8} 
                          && SPOT           & 32.75601           & 32.71165         & $\pm$ 0.18588                & 0.0339                 & 10.8925              \\ \cline{2-8}
&5                         & AMZN           & 32.45834           & 32.37494         & $\pm$ 0.14476                & 0.0410                 & 31.3312              \\ \cline{3-8} 
                          && SPOT           & 33.06409           & 33.05937         & $\pm$ 0.18526                & 0.0388                 & 12.5131              \\ \hline\hline
130 & 2                         & AMZN           & 2.79337           & 2.80912         & $\pm$ 0.05916                & 0.0360                 & 23.5048              \\ \cline{3-8} 
                          && SPOT           & 5.02174           & 5.09682         & $\pm$ 0.09945                & 0.02997                 & 10.9041              \\ \cline{2-8} 
 & 5                         & AMZN           & 2.89101           & 2.86476         & $\pm$ 0.05876                & 0.0376                 & 31.4182              \\ \cline{3-8} 
                          && SPOT           & 5.21017           & 5.16599         & $\pm$ 0.09778                & 0.04298                 & 12.4189              \\ \hline
\end{tabular}
\caption{Comparison between PROJ and Monte-Carlo method for Asian call option. \\$S_0 = 100$, $K=70$ or $130$, $r=0.05$, $T=2$.}
\label{tab:comp-asian}
\end{table}

\begin{table}[h!]
\centering
\begin{tabular}{|p{1cm}|p{1.5cm}|p{1.5cm}|p{1.5cm}|p{2cm}|p{2.5cm}|p{1.9cm}|p{1.8cm}|}
\hline
\textbf{Strike} &
\textbf{Barrier} & \textbf{Ticker} & \textbf{PROJ} & \textbf{MC} & \textbf{Confidence MC (95\%)} & \textbf{Time PROJ (s)} & \textbf{Time MC (s)} \\ \hline\hline

 80 & 65                        & AMZN           & 23.75634           & 23.78660         & $\pm$ 0.16360                & 0.9641                  & 23.9293              \\ \cline{3-8} 
                          && SPOT           & 25.90342           & 25.78946         & $\pm$ 0.21257                & 0.8956                  & 10.8809              \\ \cline{2-8}
  & 70                        & AMZN           & 23.53713           & 23.56168         & $\pm$ 0.16431                & 0.8675                  & 22.8301              \\ \cline{3-8} 
                          && SPOT           & 25.38192           & 25.36457         & $\pm$ 0.21391                & 0.8563                  & 10.7869              \\ \hline\hline
120 & 65                        & AMZN           & 7.09585           & 7.11807         & $\pm$ 0.10094                & 0.7670                 & 22.7589              \\ \cline{3-8} 
                          && SPOT           & 10.53853           & 10.42824         & $\pm$ 0.14918                & 0.8172                 & 10.9073              \\ \cline{2-8}
 & 70                        & AMZN           & 7.07874           & 7.04099         & $\pm$ 0.10074                & 0.7659                 & 22.6380              \\ \cline{3-8} 
                          && SPOT           & 10.44223           & 10.47164         & $\pm$ 0.15067                & 0.7658                 & 10.8705              \\ \hline
\end{tabular}
\caption{Comparison between PROJ and Monte-Carlo method for Barrier (down-and-out) call option. \\$S_0 = 100$, $K=80$ or $120$, $r=0.05$, $T=1$.}
\label{tab:comp-barrier}
\end{table}

\end{document}